 \documentclass[final,5p,times,twocolumn,authoryear]{elsarticle}

\usepackage[T1]{fontenc}
\usepackage[utf8]{inputenc}

\usepackage{graphicx}
\usepackage{amssymb}
\usepackage{amsmath}
\usepackage{textgreek}

\usepackage{multirow}
\usepackage{multicol}
\usepackage{fourier}
\usepackage{booktabs}
\usepackage{threeparttablex}
\usepackage{longtable}
\usepackage{xfrac}
\usepackage{booktabs}

\renewcommand{\cite}{\citep}

\usepackage{lineno}

\usepackage{subcaption}

\usepackage[colorinlistoftodos,prependcaption]{todonotes}
\usepackage{regexpatch}
\makeatletter
\xpatchcmd{\@todo}{\setkeys{todonotes}{#1}}{\setkeys{todonotes}{inline,#1}}{}{}
\makeatother

\usepackage{upgreek}

\biboptions{round}

\journal{arXiv}

\begin{document}

\begin{frontmatter}



\title{Sampling theory for spatial field sensing: Application to electro- and magnetoencephalography}



\author[nbe]{Joonas Iivanainen\corref{cor}}
\ead{joonas.iivanainen@aalto.fi}
\author[nbe]{Antti Mäkinen\corref{cor}}
\ead{antti.makinen@aalto.fi}
\author[nbe]{Rasmus Zetter}
\author[nbe]{Matti Stenroos}
\author[nbe]{Risto J.~Ilmoniemi}
\author[nbe]{Lauri Parkkonen}

\address[nbe]{Department of Neuroscience and Biomedical Engineering, Aalto University School of Science, FI-00076 Aalto, Finland}

\cortext[cor]{These authors contributed equally to this work.}

\begin{abstract}
We present a theoretical framework for analyzing spatial sampling of fields in three-dimensional space. The framework bridges Shannon's sampling and information theory to Bayesian probabilistic inference and experimental design. Based on the theory, we present an approach for optimal sampling on curved surfaces and analyze spatial sampling of EEG as well as that of on- and off-scalp MEG. Our spatial-frequency analysis of simulated measurements shows that the available spatial degrees of freedom in the electric potential are limited by the smoothing due to head tissues, while those in the magnetic field are limited by the measurement distance. On-scalp MEG would generally benefit from three times more samples than EEG or off-scalp MEG. Assuming uniform whole-head sampling and similar signal-to-noise ratios, on-scalp MEG has the highest total information content among these modalities. If the number of spatial samples is small, nonuniform sampling can be beneficial.
\end{abstract}


\end{frontmatter}



\section{Introduction}
\noindent Reconstruction of a continuous signal from discrete noisy measurements is a fundamental task in sampling. Theorems that dictate the necessary conditions for accurate signal reconstruction, such as the number of samples, are essential for guiding data acquisition (e.g., \citealt{shannon1949,pesenson2000sampling, mcewen2011novel}). The most well known of these theorems is the Shannon--Nyquist sampling theorem of bandlimited functions \cite{shannon1949, luke1999, unser2000}, which connects continuous 1D signals and their discrete representations. The theorem applies also to separable multi-dimensional signals such as two- and three-dimensional images.

The conditions for the location and count of sampling points dictated by sampling theorems
cannot always be conformed to, for example, due to insufficient prior knowledge of the signal, limited number of sensors, or restrictions in placing the sensors. If the sampling geometry is complex (such as an arbitrary surface), there may be no applicable general theorem to guide sampling and the optimal sampling positions have to be determined using methods specific to the problem at hand \cite{krause2008near}.

Here, we present a theoretical framework for analyzing spatial sampling of fields in 3D space and especially on curved surfaces. Based on the theory, we describe a method to determine optimal sampling positions. We apply the theory and the method for spatial sampling of the electric (EEG) and magnetic fields (MEG) produced by electric activity of the human brain. 


We consider the fields of interest as Gaussian random fields that can be used to represent the spatial correlations of the field \cite{abrahamsen1997review}. The random fields are used to encode prior knowledge and to make Bayesian inferences of the field. The random-field perspective is similar to kriging in geospatial sciences \cite{chiles2018} and Gaussian process regression in machine learning \cite{williams2006}. In particular, we represent the random field using a set of basis functions. The prior of the random field can be expressed conveniently using the coefficients of these basis functions.

We construct appropriate basis functions for representing the field on a curved surface. One such basis is the eigenbasis of the surface Laplace--Beltrami operator \cite{reuter2009,bronstein2017geometric} comprising spatial-frequency functions. This basis is a natural generalization of the 1D Fourier basis to a surface. These basis functions can be used to analyze the spatial degrees of freedom as well as the energy distributions of the fields. We also describe how bases that compress the field representation can be constructed.

Together with the basis-function perspective, we relate the random field to Shannon's sampling and information theory \cite{shannon1949} as well as to Bayesian experimental design \cite{lindley1956measure, chaloner1995bayesian, krause2008near}. From the theory, we also derive measures that can be used to quantify the goodness of the sampling positions. We introduce a method that maximizes Shannon's information \cite{shannon1949} for a given number of samples on the domain.

\section{Spatial sampling of electro- and magnetoencephalography}

\noindent Electroencephalography (EEG) and magnetoencephalography (MEG) are noninvasive neuroimaging techniques for measuring brain activity at millisecond time scale \cite{nunez2006electric,hamalainen1993magnetoencephalography}. EEG and MEG measure the electric and magnetic field due to neuronal currents. Adequate spatial sampling is necessary so that the discrete measurements represent the continuous field as accurately as possible so that the full detail of the field can be used to make inference of brain activity. EEG setups typically involve 16--128 electrodes placed uniformly on the subject's scalp while state-of-the-art MEG systems use around 300 superconducting quantum interference device (SQUID) sensors placed rigidly around the subject's head \cite{baillet2017magnetoencephalography}.



Spatial sampling of EEG and MEG has become a topic of significant interest again as it has been argued that the detail these techniques provide of the neuronal activity can be increased. In EEG, the number of electrodes on the scalp can be increased; high-density EEG (hdEEG) with electrode count in the range of 128--256 has attracted considerable interest to exploit the full spatial detail of the electric potential (e.g., \citealt{brodbeck2011electroencephalographic}). In MEG, in contrast to conventional liquid-helium-cooled SQUID sensors that measure the field $\sim$2 cm off the scalp, novel sensors such as optically-pumped magnetometers (OPMs; \citealt{budker2007optical}) and high-T$_\mathrm{c}$ SQUIDs \cite{faley2017high} have enabled on-scalp field sensing within millimetres from the head surface. Simulations have shown that the closer proximity of the sensors to the brain improves spatial resolution of MEG and increases information about neuronal currents \cite{schneiderman2014information, boto_potential_2016, iivanainen_measuring_2017, riaz2017evaluation}. 


The number of sensors or, equivalently, sensor spacing in EEG has been extensively studied both with simulations and experiments, while that of MEG has received less attention. Srinivasan and colleagues (\citeyear{srinivasan1998}) argued that to characterize the full range of spatial detail available for EEG, at least 128 sensors are needed (electrode spacing $\sim$3 cm). Simulations have shown that the spatial low-pass filtering by the layered conductivity structure of the head limits useful sensor spacing in EEG \cite{srinivasan1996}. More recently, finite-element modeling in a realistic head model suggested a minimum electrode spacing of 50--59 mm determined by the distance at which the potential decreased to 10$\%$ of its peak \cite{slutzky2010optimal}. Experiments by Freeman and colleagues (\citeyear{freeman2003spatial}) suggested that electrode spacing as small as 5--8 mm could be beneficial, if high spatial frequencies contain behaviorally relevant information above the noise level; however, no evidence of such high-spatial-frequency content was given. Additionally, several studies have suggested that hdEEG (electrode count 128--256 or even more) would be beneficial (e.g., \citealt{brodbeck2011electroencephalographic,petrov2014ultra,grover2016,hedrich2017comparison,robinson2017very}). These studies generally assume that the noise level does not depend on the electrode count. Thus it remains uncertain whether similar benefits would be obtained simply by making larger electrodes or by improving electrode contacts by other means.


In the case of MEG, analytical calculations in half-infinite homogeneous volume conductors suggest that the sensor spacing should be approximately equal to the distance of the sensors to the brain \cite{ahonen1993}. Whether volume conduction affects the number of spatial degrees of freedom in MEG in a realistic geometry is not known as volume conduction has been studied mainly from the perspective of head model accuracy (e.g., \citealt{stenroos2014comparison}). In many MEG studies, different sensor arrays have been compared without formulating the comparison as a sampling problem \cite{wilson2007comparison,nenonen2004total,boto_potential_2016, iivanainen_measuring_2017, riaz2017evaluation}.



Previous studies have focused on sensing strategies where the sensors cover the entire scalp uniformly. However, the field can also be sampled nonuniformly. Additionally, in some applications, such as brain--computer interfaces and localization of epilectic foci, only specific parts of the brain may be of interest. In those applications, targeted sensor arrays that extract information of local cortical activity would be of value: one could reduce the sensor count by placing sensors such that the spatial resolution and sensitivity to the cortical area of interest are maximized.




\section{Theory}

\noindent 

\noindent We begin our theoretical considerations by first reviewing the Shannon--Nyquist sampling theorem and subsequently outline the same ideas in a more general context. We then introduce Gaussian random fields, generalize the standard spatial-frequency basis to curved surfaces and describe its relation to sampling. We conclude this section by introducing the developed concepts to bioelectromagnetism and by proposing strategies for designing sampling schemes for EEG and MEG.

\subsection{Shannon--Nyquist sampling theorem}
\label{sec:shannon}
\noindent To help understand the concepts developed in later sections of this work, we review the Shannon--Nyquist or Whittaker--Kotel'nikov--Shannon \cite{jerri1977shannon} theorem, following the presentation by Jerri (\citeyear{jerri1977shannon}). The theorem states that a \textit{bandlimited} signal $x(t)$ can be uniquely reconstructed from uniformly spaced time samples if the sampling frequency is larger than twice the highest frequency ($B$) of the signal. The spectral density $\hat{x}(\omega)$ of $B$-bandlimited signal $x(t)$ is zero when $|\omega| > 2\pi B$ and the signal can be represented as
\begin{equation}
    x(t) = \frac{1}{2\pi}\int_{-\infty}^{\infty} \hat{x}(\omega) e^{-i \omega t} \mathrm{d}\omega = \frac{1}{2\pi} \int_{-2\pi B}^{2\pi B} \hat{x}(\omega) e^{-i \omega t} \mathrm{d}\omega,
    \label{eq:fourier_rep}
\end{equation}
 i.e., $x(t)$ is a linear combination of Fourier basis functions $e^{-i \omega t}$ weighted by a continuum of coefficients $ \hat{x}(\omega)$, $\omega \in [-2\pi B, 2\pi B]$.

Because of the band limit, the coefficients $\hat{x}(\omega)$ can be calculated as a series from uniformly-spaced samples $x_k = x(t=k/2B)$, where $k$ is an integer:
\begin{equation}
     \hat{x}(\omega) = \frac{1}{2B} \sum_{k = -\infty}^\infty  x_k e^{i k \omega / 2 B}.
\end{equation}
Substituting the coefficients back into Eq.~\eqref{eq:fourier_rep} reconstructs the continuous signal from the samples and gives the Whittaker--Shannon interpolation formula:
\begin{equation}
    x(t) = \sum_{k = -\infty}^\infty x_k \; \mathrm{sinc}\left( 2Bt - k \right),
    \label{eq:shannon_interpolation}
\end{equation}
where $\mathrm{sinc}(x) = \sin(\pi x)/(\pi x)$. The highlights are: 1) assumption about bandlimited signal, 2) a stable way to calculate the basis-function coefficients from the samples, and 3) an interpolation formula based on these coefficients. 



\subsection{Sampling and reconstruction of functions with finite support}
\label{sec:sampling_finite}

\noindent In this subsection, we introduce a more general treatment for sampling and reconstructing functions in 3D space. We assume that the functions have a finite support, i.e., they can be represented as finite linear combinations of some basis functions.

We start by assuming that our signal of interest, which is now a field in 3D space, is supported by up to $B$ coefficients in basis $\{\psi_m\}$
\begin{equation}
    f( \vec r) = \sum_{m=1}^\infty a_m \psi_m( \vec r) =  \sum_{m=1}^B a_m \psi_m( \vec r) = \boldsymbol{\psi}(\vec r)^\top\mathbf{a},
    \label{eq:coef_basis}
\end{equation}
where $\boldsymbol{\psi}(\vec r)$ and $\mathbf{a}$ are $(B\times 1)$ column vectors with elements $\boldsymbol{\psi}(\vec r)[m] = \psi_m( \vec r)$ and $a_m$, respectively. For a collection of $N$ noisy samples $ \{ y_n \}$ of the continuous field $f( \vec r)$ at locations $\vec r_n$, we let the ($N\times 1$) vector $\mathbf{y}$ denote the samples:
\begin{equation}
    \mathbf{y}  
    = \begin{bmatrix} 
    \boldsymbol{\psi}({\vec r_1})^\top\\
    \vdots \\
    \boldsymbol{\psi}({\vec r_N})^\top \\
    \end{bmatrix}
    \mathbf{a} 
    + \boldsymbol{\epsilon}
    = \boldsymbol{\Psi} \mathbf{a} + \boldsymbol{\epsilon},
    \label{eq:matrixmeas}
\end{equation}
where $\boldsymbol{\Psi}$ is an $(N \times B)$ matrix with the column vectors $\boldsymbol{\psi}$ at $\vec r_n$ on rows, i.e., $\boldsymbol{\Psi}[n,m] = \psi_m ( \vec r_n)$, and $\boldsymbol{\epsilon} \sim \mathcal{N}(0, \sigma^2\mathbf{I})$ is white measurement noise. 

The continuous field $f(\vec r)$ can be reconstructed from the samples $\mathbf{y}$ by estimating the coefficients $\mathbf{a}$ and using Eq.~\eqref{eq:coef_basis} to interpolate the function for any point $\vec r$ in the domain. Assuming $N>B$, we estimate the coefficients $\mathbf{a}$ by minimizing the sum of squared errors:
\begin{equation}
\hat{\mathbf{a}}=\underset{\mathbf{a}}{\text{argmin}}\qquad  \|\mathbf{y} - \boldsymbol{\Psi} \mathbf{a} \|^2,
\end{equation}
yielding the standard least-squares solution
\begin{equation}
    \hat{\mathbf{a}} = (\boldsymbol{\Psi}^\top\boldsymbol{\Psi})^{-1}\boldsymbol{\Psi}^\top \mathbf{y} = \mathbf{G}\mathbf{y}.
    \label{eq:ord_least_squares}
\end{equation}
Using Eq.~\eqref{eq:coef_basis} and Eq.~\eqref{eq:ord_least_squares}, the field can be reconstructed as
\begin{equation}
    \hat{f}(\vec r) = \boldsymbol{\psi}(\vec r)^\top \mathbf{G}\mathbf{y} = \sum_n \left(\sum_m \mathbf{G}[m,n]\psi_m(\vec r)\right) y_n.
    \label{eq:ord_least_squares_interp}
\end{equation}
To summarize, we assumed a function with a finite support, found a way to compute the coefficients from the samples and obtained an interpolation formula that reconstructs $f$ from the measurements. These steps are analogous to those in Sec. \ref{sec:shannon}. However, in contrast to this analysis, in the Shannon--Nyquist theorem the sampling locations are specified. To analyze the role of the sampling locations, we examine the problem further.


Inserting Eq.~\eqref{eq:matrixmeas} into Eq.~\eqref{eq:ord_least_squares}, we get $\hat{\mathbf{a}} = \mathbf{a} + \mathbf{G}\boldsymbol{\epsilon}$. The matrix $\mathbf{G}$ maps the noise into the estimates; the  effect of noise can be quantified by calculating the covariance of the coefficient error
\begin{equation}
    \operatorname{Cov}[\hat{\mathbf{a}} - \mathbf{a}] = \mathbf{G}\operatorname{Cov}[\boldsymbol{\epsilon}]\mathbf{G}^\top
    = \sigma^2\mathbf{G}\mathbf{G}^\top
    = \sigma^2(\boldsymbol{\Psi}^\top\boldsymbol{\Psi})^{-1}.
    \label{eq:coef_covariance}
\end{equation}
Assuming the basis $\{\psi_m\}$ orthonormal over the sampling domain, the covariance can be used to calculate the expected field reconstruction error as measured by the $L^2$ norm
\begin{equation}
\begin{split}
\operatorname{E} \|f  - \hat{f} \|_{L^2}^2
&= \operatorname{E} \| \sum_{m=0}^M (a_m - \hat{a}_m) \psi_m \|_{L^2}^2
\\ &= \operatorname{E} \| \mathbf{a} - \hat{\mathbf{a}}\|^2 =
\operatorname{Tr}[\operatorname{Cov}(\mathbf{a} - \hat{\mathbf{a}})]
\\ &=  \sigma^2\operatorname{Tr}\{ (\boldsymbol{\Psi}^\top\boldsymbol{\Psi})^{-1} \}.  
\label{eq:min_over_space}
\end{split}
\end{equation}
If we wish to minimize the expected field reconstruction error assuming that the measurement noise is independent of the sampling locations $R = \{\vec{r}_n\}$, then the trace of $(\boldsymbol{\Psi}^\top\boldsymbol{\Psi})^{-1}$ must be minimized; this also minimizes the overall error in the coefficients.

We made several assumptions and simplifications in the above analysis. First, the noise was assumed white. Second, the field was assumed to have a finite support. If the field is not strictly finitely supported, we get reconstruction error from truncation as well as from aliasing if the data due to components outside the assumed support are not in the null space of $\mathbf{G}$. Third, the size of the support $B$ was assumed smaller than the number of samples. Fourth, in Eq.~\eqref{eq:min_over_space} the basis was assumed orthonormal. In the next section, we tackle the sampling problem with a more general Bayesian approach and notice that the above treatment is a special case of that.

\subsection{Random fields and Bayesian formulation of sampling and reconstruction}
\label{sec:bayes_sampling}



\noindent In addition to the measured samples, field reconstruction is based on \emph{a priori} knowledge of the field of interest and noise. In the Bayesian framework, this knowledge is cast explicitly to prior probability distributions. With the basis-function perspective, the prior can be given for the basis-function coefficients $a_m$ (Eq.~\eqref{eq:coef_basis}), which are considered random variables. We assume $a_m$ to be Gaussian, whose joint probability density is denoted as $\mathcal{N}(\mathbf{m}_a, \mathbf{K}_a)$. Here, $\mathbf{m}_a$ is the mean (vector) of $\mathbf{a}$ and $\mathbf{K}_a$ is the covariance matrix of $\mathbf{a}$ representing the prior uncertainty about the coefficients.
 
 The prior uncertainty of $\mathbf{a}$ maps to uncertainty in the field $f( \vec r) = \sum_m a_m \psi_m( \vec r)$. Since the mapping $\mathbf{a}\mapsto f(\vec{r})$ is linear, $f( \vec r)$ is Gaussian distributed, i.e., $f( \vec r)$ is a \emph{Gaussian random field} determined by its mean field
  \begin{equation}
    \mu_f(\vec{r}) = \operatorname{E}(f(\vec{r})) = \sum_k    \boldsymbol{\mu}_a[m] \psi_k(\vec{r}) = \boldsymbol{\psi} (\vec{r})^\top \boldsymbol{\mu}_a
  \end{equation}
  and \textit{covariance kernel}
  \begin{equation}
    \begin{split}
      K_f(\vec{r}, \vec{r}\,') &= \operatorname{Cov}(f(\vec{r}), f(\vec{r}\,'))\\ &=\sum_{m,m'} \mathbf{K}_a[m,m']\psi_m(\vec{r}) \psi_{m'}(\vec{r}\,')  \\ &= \boldsymbol{\psi}(\vec{r})^\top \mathbf{K}_a \boldsymbol{\psi}(\vec{r}\,')
      \label{eq:kernel_def}.
    \end{split}
  \end{equation}
  The mean field represents expected values of the field based on prior knowledge and the variance $K_f(\vec{r}, \vec{r}\,'=\vec{r}\,)$ the uncertainty around the mean. Every finite collection of samples of the random field is jointly distributed as $\mathcal{N}(\boldsymbol{\mu}, \mathbf{K})$, where $\mathbf{K}$ is the \emph{sample covariance matrix} defined elementwise as $\mathbf{K}[n,n'] = K_f(\vec{r_n}, \vec{r}_{n'}) =  \boldsymbol{\psi}(\vec{r_n})^\top \mathbf{K}_a \boldsymbol{\psi}(\vec{r_{n'}})$ and $\boldsymbol{\mu}$ is the sample mean, i.e., $\boldsymbol{\mu}[n] = \mu_f(\vec{r}_n) = \boldsymbol{\psi} (\vec{r_n})^\top \boldsymbol{\mu}_a$.

\begin{figure*}[!t]
\centering
\includegraphics[width=1.0\linewidth]{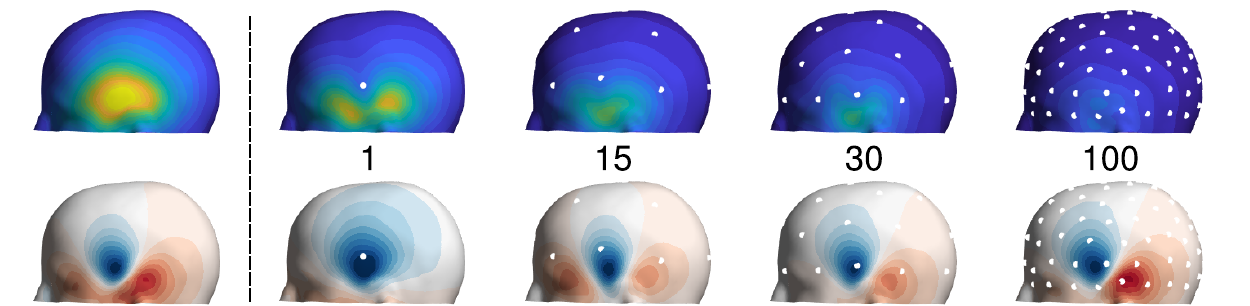}
\caption{The effect of sampling on the posterior mean field and standard deviation of the magnetic field normal component. {\bf Top left}: Prior standard deviation of magnetic field normal component due to random Gaussian sources in the temporal lobe. {\bf Bottom left}: An example field drawn from the prior distribution, sampled in the measurements on right. {\bf Top row, right}: Standard deviation of the posterior random field $\sqrt{K_f(\vec{r},  \vec{r}\,|\, R)}$ after 1, 15 30 and 100 measurements taken uniformly on the surface with an average SNR = 1. {\bf Bottom row, right}: Posterior mean field $m_f(\vec{r}\,|\,\mathbf{y}, R)$ estimating the field shown bottom left based on the 1, 15, 30, and 100  measurements. Each measurement decreases the field variance, as shown in the top row, yielding more accurate estimate of the continuous field shown on the bottom.}
\label{fig:random_field_sampling}
\end{figure*}

As described in Eq.~\eqref{eq:matrixmeas}, field measurements are always contaminated by noise. The noise sources usually include the sensors themselves, in which case the noise is usually spatially white, but also noise fields which result in correlated or colored noise. Similar to signal, the noise fields can be modeled by covariance functions $K_n(\vec{r}, \vec{r}\,') = \boldsymbol{\psi}_n(\vec{r})^\top \mathbf{K}_n \boldsymbol{\psi}_n(\vec{r}\,')$, where $\psi_n(\vec{r})$ are the basis functions for the noise field and $\mathbf{K}_n$ is the prior covariance of the noise coefficients. White sensor noise with sample covariance matrix $\boldsymbol{\Sigma} = \sigma^2\mathbf{I}$ can be modeled as equivalent field noise with the covariance kernel $K_n(\vec{r}, \vec{r}\,') = \sigma^2 \delta(\vec{r} - \vec{r}\,')$. 

The effect of measurement is encoded in the posterior probability density of the field as illustrated in Fig.~\ref{fig:random_field_sampling}. Having a collection of (noisy) samples $\mathbf{y}$ at sampling points $R = \{\vec{r}_n\,|\,n=1...N\}$, modeling the noise as  $\boldsymbol{\epsilon} \sim \mathcal{N}(0, \boldsymbol{\Sigma})$ and assuming zero mean for $f(\vec{r})$, the posterior mean field and covariance are \cite{williams2006}
\begin{equation}
\begin{split}
    \mu_f(\vec{r}\,|\,\mathbf{y}, R) 
    =& \mathbf{k}(\vec{r})^\top(\mathbf{K} + \boldsymbol{\Sigma})^{-1}\mathbf{y} ,  \\
    K_f(\vec{r},  \vec{r}\,'\,|\, R) 
    =& K_f(\vec{r}, \vec{r}\,') - \mathbf{k}(\vec{r})^\top(\mathbf{K} + \boldsymbol{\Sigma})^{-1}\mathbf{k}(\vec{r}\,'),
    \label{eq:posterior_bayes}
\end{split}
\end{equation}
where $\mathbf{k}(\vec{r})$ is the covariance between the field at the sample points and the field at the point $\vec{r}$, defined as $\mathbf{k}(\vec{r})[n] = K_f(\vec{r}, \vec{r}_n) = \boldsymbol{\psi}(\vec{r})^\top \mathbf{K}_a \boldsymbol{\psi}(\vec{r_n})$. 

Based on the measurements $\mathbf{y}$, the posterior mean $\mu_f(\vec{r}\,|\,\mathbf{y}, R)$ 
estimates the expected value of the field $f(\vec{r})$ 
yielding a reconstruction from the samples in the same way as Eqs.~\eqref{eq:shannon_interpolation} and \eqref{eq:ord_least_squares_interp}.
Specifically, the estimate is linear in $\mathbf{y}$, and we can interpret the functions in the row vector $\mathbf{k}(\vec{r})^\top(\mathbf{K} + \boldsymbol{\Sigma})^{-1}$ as interpolation functions for measurements $\mathbf{y}$. The posterior covariance $K_f(\vec{r},  \vec{r}\,'\,|\, R)$ describes the field uncertainty after the measurements at sampling locations $R$. As the matrix $(\mathbf{K} + \boldsymbol{\Sigma})^{-1}$ is positive-definite, for $\vec{r}'=\vec{r}$ the term subtracted from $K_f(\vec{r}, \vec{r})$ is always positive. Hence, the measurement reduces the field variance, i.e., the uncertainty of the field in all points sharing correlation with the field at the sampling locations. The noisier the sample, the more uncertainty is left.

In \ref{sec:appendix_mne}, we derive alternative formulas for the mean and covariance of $f(\vec r$)
\begin{equation}
\begin{split}
    \mu_f(\vec{r}\,|\,\mathbf{y}, R) 
     &= \boldsymbol{\psi}(\vec{r})^\top(\boldsymbol{\Psi^\top \Sigma^{-1} \Psi} + \mathbf{K}_a^{-1})^{-1}\boldsymbol{\Psi}^\top\boldsymbol{\Sigma}^{-1}\mathbf{y}  = \boldsymbol{\psi}(\vec{r})^\top \mathbf{\hat{a}},\\
    K_f(\vec{r},  \vec{r}\,'\,|\, R)
    &= \boldsymbol{\psi}(\vec{r})^{\top}( \boldsymbol{\Psi}^\top\boldsymbol{\Sigma}^{-1}\boldsymbol{\Psi}+ \mathbf{K}_a^{-1})^{-1} \boldsymbol{\psi}(\vec{r}).
    \label{eq:posteriorcoefs_bayes}
\end{split}
\end{equation}
Here, $\mathbf{\hat{a}} = (\boldsymbol{\Psi^\top \Sigma^{-1} \Psi} + \mathbf{K}_a^{-1})^{-1}\boldsymbol{\Psi}^\top\boldsymbol{\Sigma}^{-1}\mathbf{y}$ is posterior mean of the coefficients and $(\boldsymbol{\Psi}^\top\boldsymbol{\Sigma}^{-1}\boldsymbol{\Psi}+ \mathbf{K}_{a}^{-1})^{-1}$ is the associated posterior covariance matrix. In the limit of white noise and infinite SNR, i.e., when $\boldsymbol{\Sigma} = \sigma^2\mathbf{I}$, and $\sigma^2 \mathbf{K}_a^{-1}$ approaches zero, the formula reduces to the classical interpolation formula in Eq.~\eqref{eq:ord_least_squares_interp} and the coefficient covariance approaches $\sigma^2(\boldsymbol{\Psi}^\top\boldsymbol{\Psi})^{-1}$. Hence, the assumptions made in the previous section become more explicit.

In the following sections, we will outline how to construct basis functions for bioelectric potential and magnetic field on the measurement surfaces. The bases are then used to encode prior knowledge represented in the field kernels.

\subsection{Spatial-frequency basis}
\label{sec:fourier}
\noindent In MEG and EEG, we sample either a magnetic field component or electric potential on a curved 2D surface embedded in 3D. In this section, we describe how the conventional Fourier analysis can be extended to spatial-frequency (SF) analysis on such surfaces using a set of generalized SF basis functions. We also explain how this basis can be used to encode priors about low-spatial-frequency fields. 

Generally, spatial-frequency basis functions can be identified as eigenfunctions of the negative Laplace operator $-\nabla^2$. For example, on the real line the eigenfunctions are complex sinusoids $e^{-i \omega t}$ (see Sec. \ref{sec:shannon}) while on the surface of sphere they are the spherical harmonics. On a surface, the operator is usually called Laplace--Beltrami (LB) operator $\nabla_{\mathrm{LB}}^2$ (see, e.g., \citealt{pesenson2015sampling} and \citealt{reuter2009}). The generating equation for the spatial-frequency basis on a surface is the Helmholtz equation for $u$
\begin{equation}
    -\nabla_{\mathrm{LB}}^2 u = k^2 u,
    \label{eq:eigen_lb}
\end{equation}
where $k$ is a constant (units 1/m). The equation is an eigenvalue equation for $-\nabla_{\mathrm{LB}}^2$, where $k^2$ are eigenvalues and $u$ are the associated eigenfunctions that form the basis. Physically, the eigenfunctions $u$ represent the modes of standing waves on the domain and eigenvalues $k^2$ are the squared (spatial) frequencies of the waves. When the domain is bounded, the eigenvalue spectrum becomes discrete $\{k_i,u_i\}$. The functions $u_i$ form an orthonormal basis of square-integrable functions $L^2(S)$ on the surface $S$ \cite{pesenson2015sampling}:
\begin{equation}
    \langle u_i, u_j  \rangle = \int_S u_i(\vec r) u_j(\vec r) \mathrm{d}S = \delta_{ij},
    \label{eq:innerproduct}
\end{equation}
where $\delta_{ij}$ is the Kronecker delta. The eigenfunctions can be chosen real as the operator is real and symmetric.

\begin{figure*}[tbp]
\centering
\includegraphics[width=\linewidth]{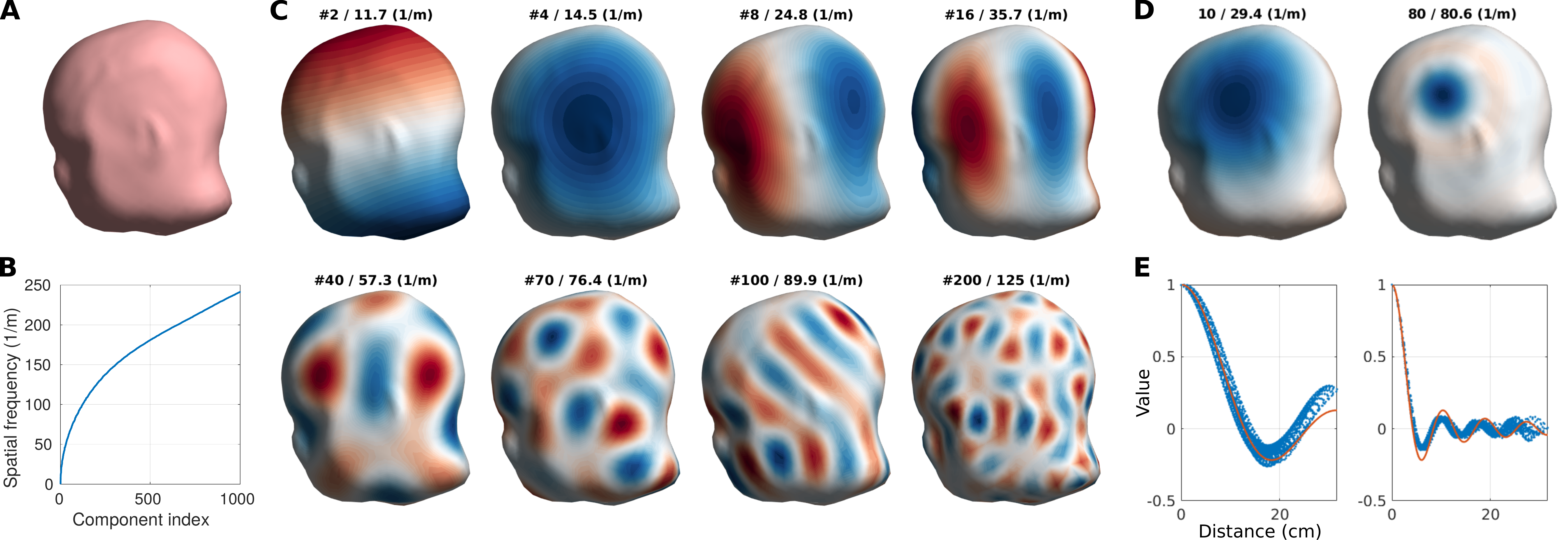}
\caption{Illustration of the eigenbasis of the negative Laplace--Beltrami (LB) operator, i.e., the spatial-frequency basis. {\bf A}: The surface on which the spatial-frequency basis was calculated. {\bf B}: Eigenvalue spectra of the LB operator, i.e., the mapping of the eigenfunction index to the spatial frequency of the eigenfunction. {\bf C}: Examples of the basis functions. Indices and spatial frequencies of the functions are shown. {\bf D}: Visualization of covariance kernels constructed with different number of spatial-frequency components for one point in the head surface. The number of components and the maximum spatial frequency in the prior are shown. {\bf E}: The normalized values (dots) of the kernel as a function of distance along the surface. Red solid lines show sinc-function fits.}
\label{fig:lbfigure}
\end{figure*}

To compute the basis in practice, the Helmholtz equation is discretized by approximating the eigenfunctions as piecewise linear on a triangle mesh that represents the surface \cite{reuter2009}
\begin{equation}
\label{eq:disclb}
    \mathbf{A}\mathbf{u}_i = k^2_i \mathbf{M u}_i,
\end{equation}
where $\mathbf{u}_i$ contains the nodal values for the $i^\mathrm{th}$ eigenfunction, $\mathbf{M}$ is a matrix that takes account the overlap in the piecewise-linear functions, and $\mathbf{A}$ is the LB matrix given by the formula \cite{macneal1949solution,jacobson2013algorithms, graichen2015sphara}
\begin{equation}
A_{i,j} =
\begin{cases}
-\frac{1}{2}(\cot{\alpha_{i,j}} + \cot{\beta_{i,j}})\quad (i,j) \text{ is an edge}  \\
- \sum_{i\neq j} A_{i,j}\quad i=j \\
\end{cases}
\end{equation}
where $\alpha_{i,j}$ and $\beta_{i,j}$ are the triangle angles opposing edge $(i,j)$. On the triangle mesh, the $L^2$ inner product can be discretized as
\begin{equation}
\label{eq:discinner}
    \langle f, g  \rangle = \int_S f(\vec r) g(\vec r)\mathrm{d}S = \mathbf{f}^\top  \mathbf{M} \mathbf{g},
\end{equation}
where $\mathbf{f}$ and $\mathbf{g}$ correspond to the coefficients of the discretizations of $f(\vec{r})$ and $g(\vec{r})$. Figure \ref{fig:lbfigure} illustrates the eigenbasis of the negative LB operator on a head surface represented as a triangle mesh.

Within the random-field context, the spatial-frequency basis can be used to represent the field kernel as $K_f(\vec{r}, \vec{r}\,')  = \mathbf{u}(\vec{r})^\top \mathbf{K_{\mathrm{SF}}} \mathbf{u}(\vec{r}\,')$, where the prior is encoded in the covariance matrix $\mathbf{K_{\mathrm{SF}}}$ and $\mathbf{u}(\vec{r})[i] = u_i(\vec{r})$. To construct the prior, we need the measurement surface and an assumption of a certain degree of smoothness. Figure \ref{fig:lbfigure}D illustrates kernels $K_f(\vec{r}_i, \vec{r})$ as functions of $\vec{r}$ constructed for different bandlimited priors (diagonal $\mathbf{K_{\mathrm{SF}}}$ with unit variance) for point $\vec{r}_i$ on a surface. For a more theoretical treatment of kernel representation with Laplacian eigenfunctions in the context of Gaussian process regression, see the work by Solin and S\"arkk\"a (\citeyear{solin2014hilbert}).

\subsection{Kernels for bioelectric potential and magnetic field}
\label{sec:bioemsampling}
\noindent The quasi-static electric potential or magnetic field component $b(\vec{r})$ is generated by distributed source activity $\vec{q}(\vec{r}_\mathrm{s})$ inside the brain and can be written in terms of Green's function $\vec{\mathcal{L}}(\vec{r},\vec{r}_\mathrm{s})$ as \cite{hamalainen1993magnetoencephalography, nunez2006electric}
\begin{equation}
     b(\vec{r}) = \int \vec{q}(\vec{r}_\mathrm{s}) \cdot \vec{\mathcal{L}}(\vec{r},\vec{r}_\mathrm{s}) dV_\mathrm{s}.
\end{equation}
For fixed $\vec r$ and varying $\vec r_\mathrm{s}$, $\vec{\mathcal{L}}$ is called the lead field; it characterizes how different source locations contribute to the field at $\vec{r}$. Correspondingly, for fixed $\vec r_\mathrm{s}$ and varying $\vec r$, $\vec{\mathcal{L}}$ is called the field topography, i.e., the field generated by a point source at $\vec{r}_\mathrm{s}$.

For notational convenience and practical implementation, we discretize the neural source distribution $\vec{q}(\vec{r}_\mathrm{s})$ into a set of primary current dipoles $\vec q_i(\vec r_i)=q_i \hat{q}_i, i=1,\ldots,N$, where $q_i$ is the amplitude of the source and $\hat{q}_i$ its direction. In consequence, Green's function discretizes to a $(N\times 1)$ lead field vector $\mathbf{l}(\vec{r})$ that contains the contribution of all elementary sources to the field at $\vec r$: $\mathbf{l}(\vec{r})[i] = t_i(\vec{r}) = \vec{\mathcal{L}}(\vec{r},\vec{r}_i)\cdot \hat{q}_i$. On the other hand, the $i$th element of the vector $\mathbf{l}(\vec{r})$ is the topography $t_i(\vec{r})$ of the $i$th source. Further, when the field is also discretized (sampled), the forward operator $\vec{\mathcal{L}}$ reduces to a matrix $\mathbf{L}$, and the field samples are given by \cite{hamalainen1993magnetoencephalography}
 \begin{equation}
     \mathbf{b} = \mathbf{L} \mathbf{q}.
 \end{equation}

Considering the source amplitudes $q_i$ as Gaussian random variables \cite{demunck1992random}, the field kernel can be written as
\begin{equation}
\label{eq:fieldsourcecovar}
    K(\vec{r},\vec{r}\,') = \mathbf{l}(\vec{r})^\top \mathbf{K}_q \mathbf{l}(\vec{r}\,') = \sum_{i,j}  \mathbf{K}_q[i,j] t_i(\vec{r})t_{j}(\vec{r}\,'),
\end{equation}
where $\mathbf{K}_q$ is the $(N\times N)$ source covariance matrix. The topographies $t_i(\vec{r})$ can be now considered as the basis functions used to assemble the kernel as in Eq.~\eqref{eq:kernel_def} and $q_i$ the associated random coefficients. For a set of sampling points $R$, the field kernel reduces to the covariance matrix $\mathbf{K} = \mathbf{L}\mathbf{K}_q\mathbf{L}^\top$.

The field kernel can be rewritten as an eigendecomposition \cite{mercer1909functions}
\begin{equation}
\label{eq:mercer}
K(\vec{r},\vec{r}\,') =  \mathbf{v}(\vec{r})^\top\mathbf{D}\mathbf{v}(\vec{r}\,') = \sum_m d_m^2 v_m(\vec{r})v_m(\vec{r}\,'),    
\end{equation}
were $v_m(\vec{r})$ form an orthonormal basis on the measurement surface, the \textit{field eigenbasis}, and $\mathbf{D}$ is a diagonal matrix of $d_m^2$. This eigenbasis enables us to express the random field as a linear combination $f(\vec{r}) = \sum_m z_m v_m(\vec{r}) $, where $z_m = \int_S f(\vec{r})v_m(\vec{r}) dS$ are independent random variables with variance $d_m^2$ \cite{loeve1978probability, stark1986probability}. 



\begin{figure*}[tbp]
\centering
\includegraphics[width=\linewidth]{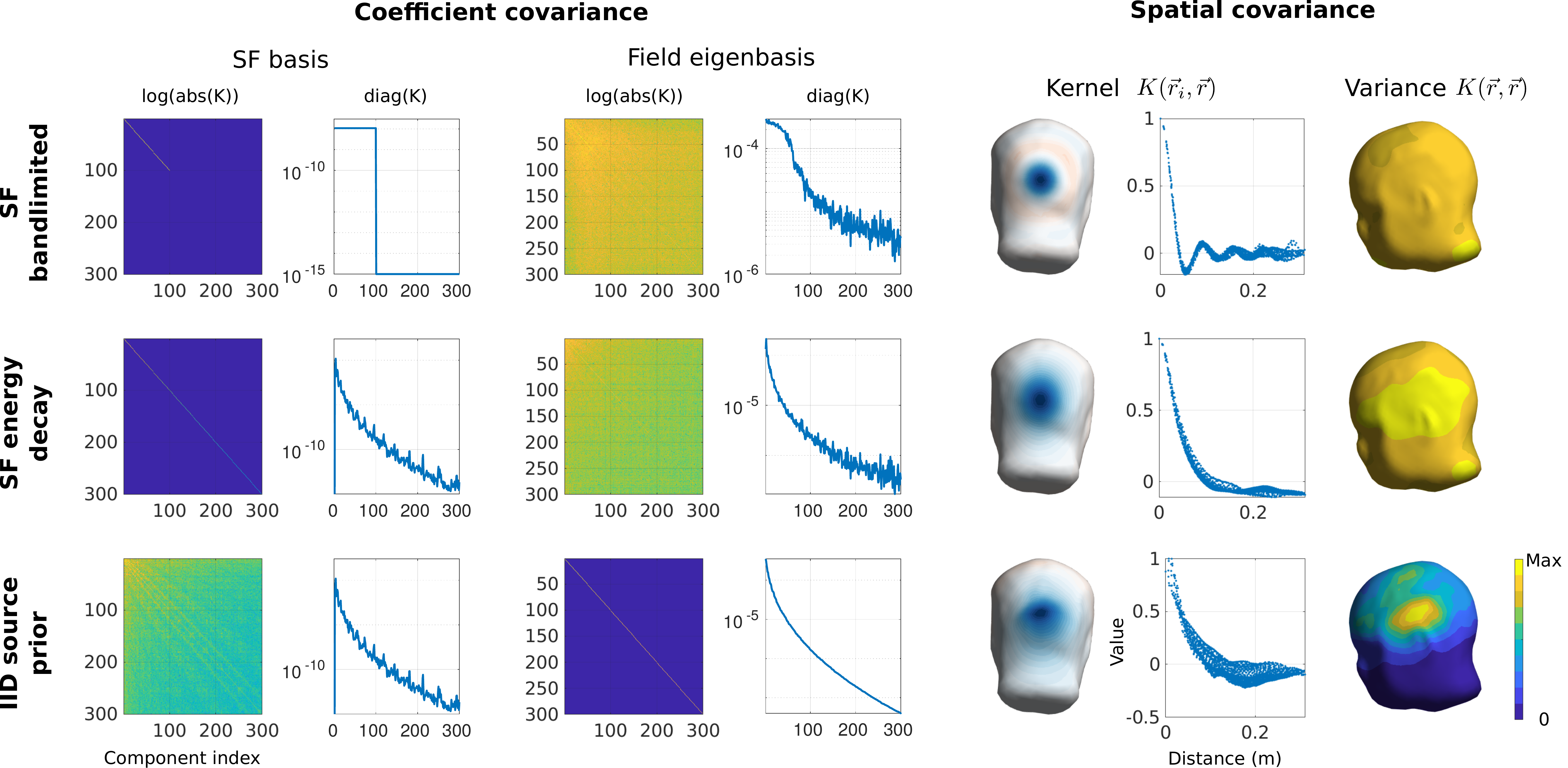}
\caption{Covariances of three field priors (SF bandlimited: bandlimited in spatial-frequency basis; SF energy decay: energy roll off in SF basis; IID source prior: identically and independently distributed sources inside the head generating the field). {\bf Left}: Prior coefficient covariance in SF basis and field eigenbasis. The covariance matrix and its diagonal are shown. {\bf Right}: Spatial covariance. The spatial kernel and its decay are shown for one position on the measurement surface. The rightmost column shows the spatial variance.}
\label{fig:kernel}
\end{figure*}

Altogether, the prior knowledge of the field behaviour can be encoded in coefficient covariances in different bases. Next, we will examine three different priors relevant to bioelectric potential and magnetic field. The first prior assumes that the field is bandlimited in the spatial-frequency basis (SF bandlimited). The second prior assumes that the field energy decays as a function of spatial frequency (SF energy decay). The third prior assumes identically and independently distributed neural sources generating the field (IID source prior), i.e., diagonal source covariance $\mathbf{K}_q = q^2 \mathbf{I}$. Fig. \ref{fig:kernel} shows the properties of these priors both in basis-function (SF basis and field eigenbasis) and spatial domains. The bandlimited SF prior (diagonal covariance with the same variance up to $k_B$) results in sinc-like spatial kernels and uniform spatial variance (Secs. \ref{sec:shannon} and \ref{sec:fourier}). A more physical prior (SF energy decay) results in smoother, symmetric spatial kernels and nearly uniform spatial variance. Both of these priors are non-diagonal in the field eigenbasis. IID source prior requires the knowledge of the measurement surface as well as a model for the sources and their fields. This prior produces non-uniform spatial variance and asymmetric spatial kernels and is non-diagonal in SF basis.

\subsection{Topography energies and implications to sampling}
\label{sec:topoenergy}
\noindent In this subsection, we present the analysis of topographies both in the spatial-frequency (SF) basis and in the field eigenbasis on the measurement surface $S$. The SF basis can be used to analyze the spatial-frequency content of the topographies. The topographies are square-integrable and therefore any topography $t(\vec{r})$ can be represented as a linear combination of the SF basis functions $t(\vec{r}) = \sum_{m=1}^\infty a_m u_m(\vec r)$ (Sec. \ref{sec:fourier}). The coefficients can be calculated as projections of $t(\vec{r})$ on $\{u_m\}$ using the $L^2$ inner product
\begin{equation}
    a_m =  \langle t, u_m  \rangle = \int_S t(\vec r) u_m(\vec r) \mathrm{d}S.
    \label{eq:innerproduct}
\end{equation}
The squared coefficients $a_m^2$ comprise the \textit{energy spectrum} of the topography. As the basis is orthonormal, the topography energy is, according to Parseval's theorem, equal in spatial and frequency domains
\begin{equation}
    ||t||_{L^2}^2 = \langle t, t  \rangle =  \int_S |t(\vec r)|^2 \mathrm{d}S = \sum_{m=1}^\infty a_m^2.
    \label{eq:parseval}
\end{equation}
For analyzing the energy content of the topographies, we define the cumulative energy in the $M$ first components as
\begin{equation}
\label{eq:cumen}
    E(M) = \sum_{m=1}^M a_m^2.
\end{equation}


Although most of the energy of the topographies lies in low spatial frequencies, the topographies are not strictly bandlimited. To analyze the reconstruction error due to assuming the topographies $B$-bandlimited, we represent them as 
\begin{equation}
\begin{split}
    t(\vec r) &= \sum_{m=1}^B a_m u_m(\vec r) + \sum_{m=B+1}^\infty a_m u_m(\vec r) \\ &= \mathbf{u}(\vec r)^\top\mathbf{a} + \mathbf{u}_\mathrm{r}(\vec r)^\top\mathbf{a}_\mathrm{r}, 
\end{split}
\end{equation} 
where the latter sum is the residual of the topography outside the band and $\mathbf{a}$ contains the associated coefficients for $m\leq B$ and $\mathbf{a}_\mathrm{r}$ for $m > B$. The energy in the first $B$ components is given by Eq. \eqref{eq:cumen}, while the residual energy outside the band is given by orthogonality
\begin{equation}
\label{eq:truncerr}
    \|\mathbf{a}_\mathrm{r}\|^2 = ||t||_{L^2}^2 - E(B).
\end{equation}

Sampling the field representation as in Sec. \ref{sec:sampling_finite}, we get
\begin{equation}
    \mathbf{y} = \boldsymbol{\Psi}\mathbf{a} + \boldsymbol{\Psi}_\mathrm{r}\mathbf{a}_\mathrm{r} + \boldsymbol{\epsilon},
\end{equation}
where $\boldsymbol{\Psi}[n,m] = u_m(\vec{r}_n)$ for $m\leq B$ and $\boldsymbol{\Psi}_\mathrm{r}[n,m] = u_m(\vec{r}_n)$ for $m > B$. Estimating the coefficients using Eq.~\eqref{eq:ord_least_squares} and the (incorrect) assumption of bandlimited energy, we get
\begin{equation}
    \hat{\mathbf{a}} = \mathbf{G}( \boldsymbol{\Psi}\mathbf{a} +\boldsymbol{\Psi}_\mathrm{r}\mathbf{a}_\mathrm{r} + \boldsymbol{\epsilon}) 
    =  \mathbf{a} + \mathbf{G}(\boldsymbol{\Psi}_\mathrm{r}\mathbf{a}_\mathrm{r} + \boldsymbol{\epsilon}) = \mathbf{a} + \Delta\mathbf{a}
\end{equation}
where $\Delta\mathbf{a} = \mathbf{G}(\boldsymbol{\Psi}_\mathrm{r}\mathbf{a}_\mathrm{r} + \boldsymbol{\epsilon})$ is the error in the estimated coefficients due to the residual aliasing and measurement noise. Due to the orthogonality of the subspaces below and above $B$ and the zero mean of the white noise, the topography reconstruction error evaluated using the $L^2$ norm can be expressed as
\begin{equation}
\label{eq:fieldrecoerr}
\begin{split}
    \operatorname{E}\|t - \hat{t}\|_{L^2}^2 &=  \operatorname{E}\|\mathbf{a}-\hat{\mathbf{a}}\|^2 + \|\mathbf{a}_\mathrm{r}\|^2 = \operatorname{E}\|\Delta\mathbf{a}\|^2 + \|\mathbf{a}_\mathrm{r}\|^2 \\ &= \|\mathbf{G}\boldsymbol{\Psi}_\mathrm{r}\mathbf{a}_\mathrm{r}\|^2 + \operatorname{Tr}[\sigma^2({\boldsymbol{\Psi}^{\top}\boldsymbol{\Psi}})^{-1}] + \|\mathbf{a}_\mathrm{r}\|^2 \\
    &\leq (1+C)\|\mathbf{a}_\mathrm{r}\|^2 + \operatorname{Tr}[\sigma^2({\boldsymbol{\Psi}^{\top}\boldsymbol{\Psi}})^{-1}],
\end{split}
\end{equation}
where $\|\mathbf{\mathbf{G}\boldsymbol{\Psi}_\mathrm{r}\mathbf{a}_\mathrm{r}}\|^2$ is the aliasing error, $\operatorname{Tr}[\sigma^2(\boldsymbol{\Psi}^{\top}\boldsymbol{\Psi})^{-1}]$ the error due to white noise (Eq.~\eqref{eq:min_over_space}), $\|\mathbf{a}_\mathrm{r}\|^2$ (Eq. \eqref{eq:truncerr}) is the \textit{truncation error} and $C$ is a positive constant. Apart from the measurement noise, the reconstruction error is thus bounded by the truncation error. The constant $C$ in the bound depends on the sampling pattern via the matrix $\mathbf{\mathbf{G}}\boldsymbol{\Psi}_\mathrm{r}$.

Finally, we describe the connection between energies of the individual topographies and the decomposition of variance of the topographies of random sources. A random source distribution produces a random topography $f(\vec{r})$ which can be represented in an orthonormal basis as described above. The expected energy of $f(\vec{r})$ (with zero mean) can be calculated as
\begin{equation}
\begin{split}
    \operatorname{E}(\|f\|_{L^2}^2) &= \sum_{m=1}^\infty \operatorname{E}({a_m}^2) \\
    &= \sum_{m=1}^\infty \int \int u_m(\vec r)  \operatorname{E}[f(\vec r)  f(\vec r\,')] u_m(\vec r\,') dS' dS \\
    &= \sum_{m=1}^\infty \int \int u_m(\vec r)  K(\vec r,  \vec{r}\,') u_m(\vec r\,') dS'  dS,
\end{split}
\end{equation}
where the double integral in the last expression can be identified as the variance of the coefficient $a_m$.

IID-distributed random sources produce a random topography with kernel $K(\vec{r}, \vec{r}\,') = q^2 \mathbf{l}(\vec{r})^\top \mathbf{l}(\vec{r}\,') = q^2\sum_m t_m(\vec{r}) t_m(\vec{r}\,')$ and the coefficient variance can be calculated as
\begin{equation}
\begin{split}
    \operatorname{E}({a_m}^2) &= \int\int u_m(\vec{r}) K(\vec{r}, \vec{r}\,') u_m(\vec{r}\,') dS' dS 
    \\ &= q^2 \sum_i <t_i, v_m>^2.
    \label{eq:eigen_variance}
    \end{split}
\end{equation}
In other words, the variance projected to a single basis function can be calculated as the product of source variance $q^2$ and the sum of the squared spectral coefficients of the topographies $t_i$. The energy distribution of single topographies thus contain information about the maximum number of degrees of freedom in the field of any source distribution.

The topography energy can also be decomposed in the eigenbasis $\{v_m\}$ of the modeled kernel $K(\vec{r}, \vec{r}\,')$ (Eq.~\eqref{eq:mercer}). Since the eigenfunctions $v_m(\vec{r})$ are orthonormal, the coefficients for a single topography can be calculated with the inner product in Eq.~\eqref{eq:innerproduct} and the "energy spectrum" in this basis is given by the squared coefficients $a_{i,m}^2 = <t_i, v_m>^2$. The Karhunen--Loève theorem [\citealt{stark1986probability}, section 7.6] states that the eigenfunctions provide an optimal representation of the random field in the sense that it minimizes the expected truncation error. Thus, compared to the SF basis $\{u_m\}$, on average fewer terms are needed to reach a certain truncation error.


\subsection{Optimality criteria of sampling grids}
\label{sec:optimalcriteria}
\noindent In section \ref{sec:bioemsampling}, we saw that field kernels constructed with different basis-function sets and coefficient priors have different profiles of spatial covariance. In this and the following section, we will discuss how optimal sampling-point sets $R$ can be constructed using that information. 

Generally, we want to minimize the reconstruction error of Eq.~\eqref{eq:min_over_space}. As shown in the previous sections, this problem can be re-formulated in ways that involve the inversion of matrix $\boldsymbol{\Psi}^\top\boldsymbol{\Psi}$ or $\mathbf{K} + \boldsymbol{\Sigma}$. One way to look at the optimality is then to maximize the stability of the inversion. The size of the inverted matrix should be thus as "large" as possible. Matrix size can be generally quantified in multiple ways, the different notions of size giving different optimality criteria in the context of \textit{(Bayesian) experimental design} \cite{krause2008near, chaloner1995bayesian}.


In probabilistic sense, we can interpret the criteria above as minimization of coefficient variance which is called \textit{(Bayesian) A-optimality} \cite{krause2008near} in optimal design. Classically, the overall variance in the coefficient estimates (see Sec.~\ref{sec:sampling_finite}) can be measured by $\operatorname{Tr}[(\boldsymbol{\Psi}^\top\boldsymbol{\Psi})^{-1}]$. From the Bayesian perspective, assuming noise covariance $\boldsymbol{\Sigma}$ and prior coefficient covariance $\mathbf{K}_a$, the objective function would be the posterior coefficient variance $\operatorname{Tr}[(\boldsymbol{\Psi}^\top\boldsymbol{\Sigma}^{-1}\boldsymbol{\Psi}+ \mathbf{K}_{a}^{-1})^{-1}]$ (Sec.~\ref{sec:bayes_sampling}). When studying sufficient sampling on a bounded surface as in MEG and EEG, we should, however, measure how well the sampling pattern captures the overall field variance on the surface. To optimize the measurement in this sense, we define a new objective, the fractional explained variance
\begin{equation}
    \mathrm{FEV}(R) = 1- \frac{\int K_f(\vec{r},  \vec{r}\,|\, R) dS}{\int K_f(\vec{r}, \vec{r}) dS} 
    \propto \int \mathbf{k}(\vec{r})^\top(\mathbf{K} + \boldsymbol{\Sigma})^{-1}\mathbf{k}(\vec{r})dS ,
    \label{eq:fev}
\end{equation}
which ranges from 0 (no variance explained) to 1 (all variance explained). This measure is related to \textit{I-optimality} or integrated optimality \cite{atkinson2014optimal}.

The matrix size can also be measured with the determinant; its maximization results in \textit{D-optimality} \cite{chaloner1995bayesian}. In the Bayesian setting, this leads to maximizing the total information of the measurement
\begin{equation}
    \mathrm{info}(R) = \frac{1}{2} \log_2 \frac{\det(\mathbf{K} + \boldsymbol{\Sigma})}{\det(\boldsymbol{\Sigma})} = \frac{1}{2} \log_2 \det(\tilde{\mathbf{K}} + \mathbf{I}),
    \label{eq:info}
\end{equation}
where $\tilde{\mathbf{K}} = \boldsymbol{\Sigma}^{-1/2} \mathbf{K} \boldsymbol{\Sigma}^{-1/2}$ is the \textit{whitened} covariance. Note that the random field can also be whitened already before sampling as explained below. The D-optimal $R$ minimizes the posterior entropy of the random-field coefficients \cite{sebastiani2000maximum}, i.e., maximizes the information gained, e.g., from the field sources. Previously, total information has been used in studies comparing MEG sensor arrays \cite{kemppainen1989, nenonen2004total,schneiderman2014information,iivanainen_measuring_2017, riaz2017evaluation}. The equivalence of Eq.~\eqref{eq:info} and the definition of information in these studies is shown in \ref{sec:apppendix_info}.

To assess the relative power of signal and noise, we define spatial signal-to-noise ratio $\mathrm{SNR}(\vec{r})$ as the ratio of signal $K_f(\vec{r}, \vec{r})$ and noise variances $K_n(\vec{r}, \vec{r})$
in case of uncorrelated noise. For a more general definition, we decompose the noise kernel as $K_n(\vec{r}, \vec{r}\,') =  \mathbf{w}(\vec{r})^\top\boldsymbol{\Lambda}\mathbf{w}(\vec{r}\,')$, where $w_i(\vec{r})$ are the eigenfunctions of the noise kernel and $\boldsymbol{\Lambda}$ contains the associated noise variance on its diagonal. The eigenfunctions can be utilized for composing a\textit{ whitening kernel} $W(\vec{r}, \vec{r}\,') = \mathbf{w}(\vec{r})^\top\boldsymbol{\Lambda}^{-1/2}\mathbf{w}(\vec{r}\,')$, which, when applied to the noise field $\int W(\vec{r}, \vec{r}\,') n(\vec{r}\,')dS'$, yields spatial white noise. Applying the whitener to the random field (signal), we get $\tilde{f}(\vec{r}) = \int W(\vec{r}, \vec{r}\,') f(\vec{r}\,')dS'$, and the spatial signal-to-noise ratio can be defined as 
\begin{equation}
\begin{split}
\mathrm{SNR}(\vec{r})&=K_{\tilde{f}}(\vec{r}, \vec{r}) \\ &= \int\int  W(\vec{r}, \vec{r}\,') K_{f}(\vec{r}', \vec{r}'')  W(\vec{r}, \vec{r}\,'') dS' dS'',    
\end{split}
\end{equation}
i.e., the transformed random field $\tilde{f}$ is in a sense a field of SNR. 

Finally the average measurement SNR can be calculated from the whitened covariance matrix $\tilde{\mathbf{K}}$ as \cite{iivanainen_measuring_2017}
\begin{equation}
    \mathrm{SNR}_\mathrm{meas} = \frac{\mathrm{tr} (\tilde{\mathbf{K}})} {\mathrm{tr} (\mathbf{I})}.
\end{equation}
Although SNR measures the average signal quality, it does not tell how well the sampling grid captures overall field variation over the sampling domain. With adequate spatial sampling, however, the measurement SNR converges to $\frac{1}{\int 1 dS} \int \mathrm{SNR}(\vec{r})dS$ as shown in Sec. \ref{sec:methods_sampling_algorithm}.

\subsection{Sampling grid construction}
\label{sec:sampling_grid_construction}

\noindent The developed theory suggests to incorporate prior knowledge of the signal and noise into the field kernels $\boldsymbol{\psi}_f(\vec{r})^\top\mathbf{K}_f\boldsymbol{\psi}_f(\vec{r}\,')$ and $\boldsymbol{\psi}_n(\vec{r})^\top\mathbf{K}_n\boldsymbol{\psi}_n(\vec{r}\,')$ corresponding to the prior coefficient covariances $\mathbf{K}_f$ and  $\mathbf{K}_n$ with the bases $\boldsymbol{\psi}_f(\vec{r})$ and $\boldsymbol{\psi}_f(\vec{r})$. The sampling grid $R$ can be obtained by maximizing a related objective function, for example the fractional explained variance in  Eq.~\eqref{eq:fev} or the total information in Eq.~\eqref{eq:info}.
In this work, we choose to maximize the total information. In \ref{sec:apppendix_info}, we show that this corresponds to maximizing the diagonal elements in the whitened sample covariance matrix $\tilde{\mathbf{K}}$ while simultaneously nulling the non-diagonal elements. In other words, maximizing the total information is equivalent to finding the sampling pattern with the least correlations and maximal signal over noise.

\begin{figure}
    \centering
    \includegraphics[width=1.0\linewidth]{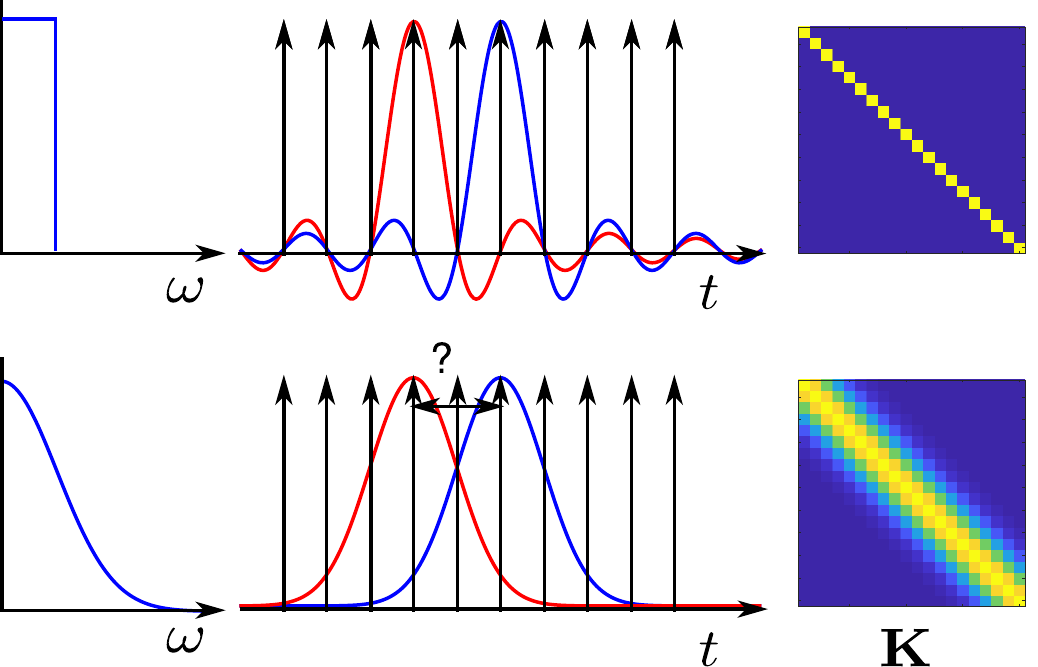}
    \caption{Illustration of covariance and sampling. {\bf Left}: Coefficient variance as a function of frequency for a strictly bandlimited process and a process with smooth variance decay. {\bf Middle}: The covariance functions and samples. The red and blue curves describe the covariance functions of two samples while the spikes are the sample positions. Equispaced samples fit the zeros of the sinc functions. For a non-sinc covariance, there is no trivial choice for the optimal sampling distance. {\bf Right}: Sample covariance matrices for the two cases, i.e., the covariance functions sampled at the equispaced locations.}
    \label{fig:sinc_gaussian}
\end{figure}

To demonstrate this briefly, we describe how the information maximization connects to the sample spacing in the Shannon--Nyquist theorem (Sec. \ref{sec:shannon}). As described before, we can treat bandlimited functions as random processes with positive uniform prior variance for the frequency coefficients up to a frequency $B$ and zero above $B$. The temporal covariance $k(t)$ is then the Fourier transformation of a boxcar function, i.e., the sinc function with zeros at equispaced intervals $1/(2B)$ illustrated in Fig.~\ref{fig:sinc_gaussian}. Considering only white noise, the sample covariance matrix $\mathbf{K}$ becomes diagonal when the sample points are chosen at the zeros of the sinc function. This sample configuration has uniform sample spacing and maximum total information.

As suggested by the circularly symmetric sinc-like covariance kernels (Figs. \ref{fig:lbfigure} and \ref{fig:kernel}), it should intuitively also hold for an SF-bandlimited field that maximizing the total information would result in a uniform measurement grid. In fact, there exists a Shannon-type sampling theorem for functions on a manifold \cite{pesenson2014multiresolution,pesenson2015sampling}, stating that an SF-bandlimited function on surface is uniquely specified by roughly uniform samples with a spacing given by $\rho = c k_B^{-1}$, where $k_B$ is the maximum spatial frequency of the function and $c$ is a positive constant that depends on the manifold.


For a general covariance kernel $K(\vec{r}, \vec{r}\,')$, finding the information-wise optimal sampling grid $R$ is, however, much more challenging (Fig. \ref{fig:sinc_gaussian}). To find an approximate solution 
to the problem of finding "as-diagonal-as-possible" covariance matrix, we reformulate the problem with the help of eigenfunctions of the continuous kernel.
First, let us consider the (whitened) kernel $K(\vec{r}, \vec{r}\,')$ with eigendecomposition $\mathbf{v}(\vec{r})^\top\mathbf{D}\mathbf{v}(\vec{r}\,')$. The decomposition can be rewritten as $K(\vec{r}, \vec{r}\,') = \mathbf{\hat{v}}(\vec{r})^\top\mathbf{\hat{v}}(\vec{r}\,')$, where $\mathbf{\hat{v}}(\vec{r}) = \mathbf{D}^{1/2}\mathbf{v}(\vec{r})$, i.e., the covariance between points $\vec{r}$ and $\vec{r}\,'$ can be calculated as a simple Euclidean inner product between $\mathbf{\hat{v}}(\vec{r})$ and $\mathbf{\hat{v}}(\vec{r}\,')$. The squared distance between the vectors is
\begin{equation}
\begin{split}
    ||\mathbf{\hat{v}}(\vec{r}) - \mathbf{\hat{v}}(\vec{r}\,')||^2 &= ||\mathbf{\hat{v}}(\vec{r})||^2 + || \mathbf{\hat{v}}(\vec{r}\,')||^2 - 2\mathbf{\hat{v}}(\vec{r})^\top\mathbf{\hat{v}}(\vec{r}\,')\\ &= K(\vec{r}, \vec{r}) + K(\vec{r}\,', \vec{r}\,') - 2K(\vec{r}, \vec{r}\,').
\end{split}
\end{equation}
Thus, maximizing pairwise distances in the eigenspace corresponds to maximizing the diagonal elements in the sample covariance matrix $\mathbf{K}$ while simultaneously minimizing the non-diagonal elements (assuming positive covariance for neighbouring samples). This is exactly what we want in the information maximization.

A solution for the optimization problem can be found with the \emph{farthest-point sampling} algorithm, which attempts to find samples in a point set by maximizing the pair-wise minimum distances between the sampled points as described by Eldar et al. (\citeyear{eldar1997farthest}) and Schl\"omer et al. (\citeyear{schlomer2011farthest}).
The algorithm works by first determining the Voronoi cells for initial sampling points, i.e., clustering all points according to their closest sample. After this, subsequently for each cell, the point farthest from all the other samples is substituted for the sampling point of the cell and the cells are updated. The iteration is continued until convergence.

In conclusion, an approximate solution for the information maximization problem can be found in two steps: (1) by embedding sampling points on the measurement surface to the eigenspace of the field kernel and (2) utilizing farthest-point sampling to maximize the minimal pairwise eigendistances between the points.

\section{Simulations}

\noindent In this section, we simulate spatial-frequency spectra of topographies and covariance kernels of source distributions in EEG and MEG. Further, we construct optimal sampling grids using different prior models and evaluate their performance. We begin the section by outlining the methods for field computation.

\subsection{Models and field computation}
\label{sec:models_and_field_computation}
\noindent We used a head model built using the example data of SimNIBS software (version 2.0) \cite{windhoff2013electric} in an earlier study \cite{stenroos2016}. The head model consists of the outer surfaces of the white matter in left and right hemispheres, gray matter, cerebellum, scalp as well as the inner and outer surfaces of the skull.


We assumed a piecewise constant isotropic conductor and computed the electric potential and magnetic field using a linear Galerkin boundary-element method with the isolated-skull approach \cite{stenroos2007matlab,stenroos2012bioelectromagnetic}. We used two source spaces consisting of point-like dipolar current sources. In the volumetric source space, the sources were set inside the volume bounded by gray-matter surface with a spacing of 4 mm, giving a total of 10 635 source positions. The minimum distance of the sources to gray-matter surface was set to 2 mm. Three orthogonally oriented sources were placed in each position to enable analysis of topographies independent of the source direction. In the cortically-constrained source space, the sources were distributed on the boundary of white and gray matter. The number of sources was 20 324 (10 225 in the left hemisphere) with an average spacing of 3.0 mm. The sources were oriented along the surface normal following the anatomical orientation of apical dendrites of pyramidal neurons.

From the head model, we constructed volume conductors with different levels of detail. The conductor (4C) used in most simulations had the following compartments: brain (inside the outer surface of gray matter) and cerebellum, cerebrospinal fluid (CSF; between the gray matter and the inner surface of the skull), skull and scalp. The conductivity of the soft tissues (brain, cerebellum and scalp) was 0.33 S/m while the conductivity of CSF was 1.79 S/m. For skull conductivity, we used the value 0.33/50 S/m. Two other conductors were constructed to analyze how volume conduction affects the fields. In the three-compartment conductor (3C) the conductivity boundaries were defined by the inner and outer skull surfaces as well as the scalp surface (brain, skull and scalp). The simplest conductor (1C) consisted of a single compartment bounded by the scalp surface.


We calculated the electric potential and the normal component of magnetic field at the nodes of triangular meshes that represented the measurement surfaces. The measurement surfaces were generated from a dense scalp mesh by cutting the mesh above a plane defined roughly by the ears and nose. The cut mesh was resampled so that it had 5404 nodes and 10421 triangles. The mesh for the electric potential (EEG) was obtained by projecting the node positions on the scalp. The on-scalp MEG surface was generated by inflating the mesh 4.5 mm away from the scalp. The more distant MEG surface (referred to as off-scalp MEG) was obtained by further inflating the mesh and by smoothing it with the function \texttt{smoothsurf} in the iso2mesh MATLAB toolbox \cite{fang2009tetrahedral}. The median distance of the nodes of the off-scalp MEG surface to the scalp was 2.4 cm (2.0--4.4 cm).


\subsection{Spatial-frequency analysis}
\label{sec:spatfreqres}

\noindent We quantified how the energies of the topographies were distributed on different spatial frequencies (SF) as explained in Sec.~\ref{sec:topoenergy}. We also examined the effect of volume conduction on these distributions. Furthermore, we estimated how many SF components were needed to obtain 99$\%$ of the energy for each topography [Eq.~\eqref{eq:cumen}] The maximum spatial frequency of those components was defined as the 99\% bandwidth of the topography.
\paragraph{Methods} 

We calculated the SF basis as described in Sec. \ref{sec:fourier}. As the measurement surfaces were not closed, a boundary condition (BC) had to be set. Here, we used the zero-Neumann BC, which sets the outwards-facing derivative of $u$ to zero. This BC was used as it accounts better for field energy near the boundary than the Dirichlet BC, which fixes the value of $u$ on the boundary. The LB operator $\mathbf{A}$ (Eq. \eqref{eq:disclb}) was computed using MATLAB function \texttt{cotmatrix} and the matrix $\mathbf{M}$ using \texttt{massmatrix}, both functions included in gptoolbox \cite{gptoolbox}. The generalized eigenvalue equation (Eq. \eqref{eq:disclb}) was solved in MATLAB for the smallest eigenvalues.

The SF coefficients were computed in the 4C model using the discretized inner product (Eq. \eqref{eq:discinner}) and were squared to obtain the energy spectrum. We used the volumetric source space and calculated the energy spectra as the sum of the three orthogonal sources in each position to analyze the topography energy spectra independent of the the source direction. The number of SF components to obtain 99$\%$ of the topography energy was computed for each modality and for each source position. The analysis was repeated for the two other volume conductors, to see how the properties of the volume conductor affect the spectra, as well as for the cortically-constrained source space. The sensor spacing in each modality corresponding to the maximum of the 99\% topography bandwidths across the source positions was estimated by generating uniform grids (Sec. \ref{sec:methods_sampling_algorithm}). 

\begin{figure*}[!t]
\centering
\includegraphics[width=\linewidth]{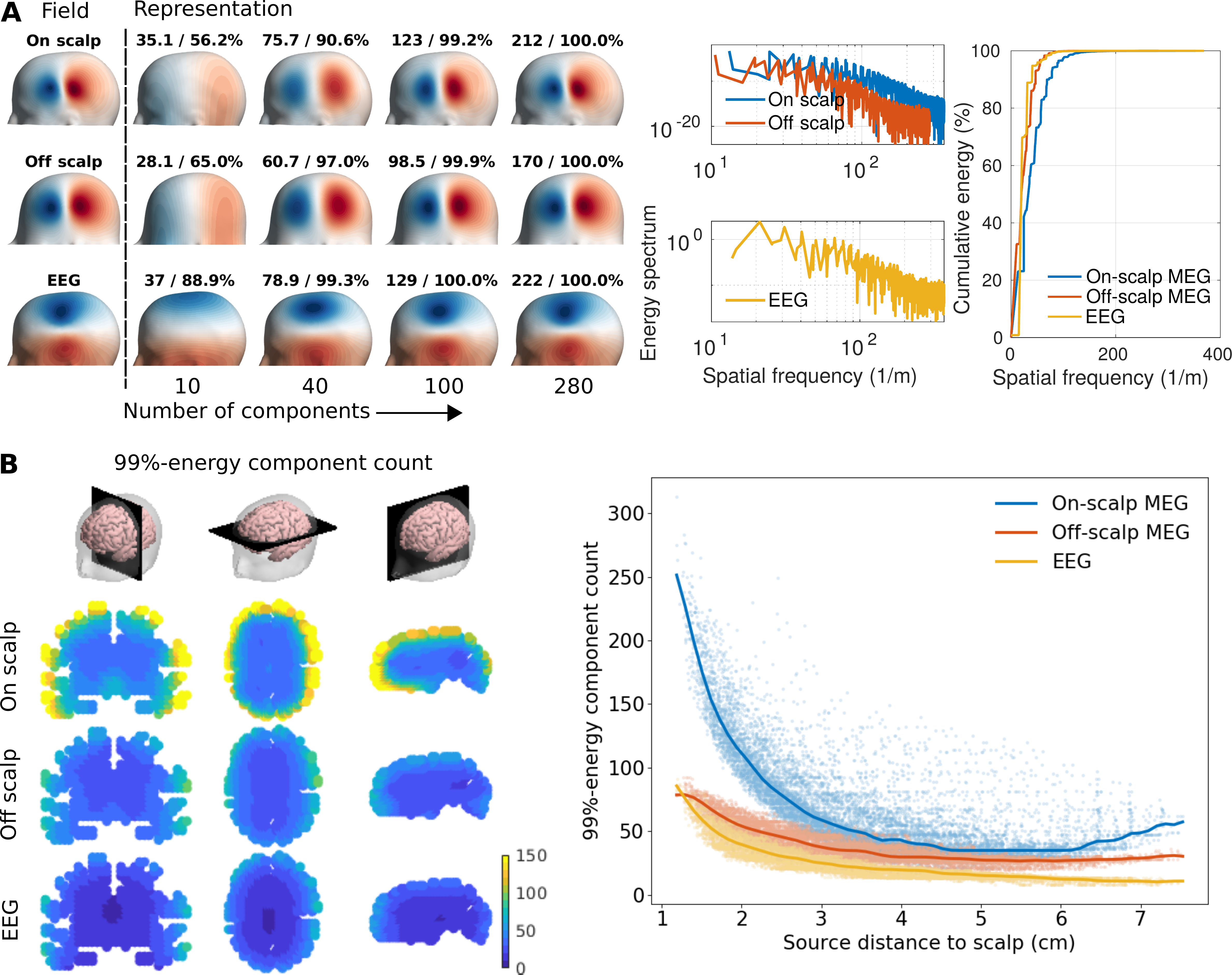}
\caption{Spatial-frequency (SF) representations of field topographies. {\bf A, left:} Representations of a single topography with increasing number of SF components. The maximum spatial frequency (1/m) and the cumulative energy of the representation are shown. {\bf A, right:} Energy spectrum and cumulative energy of the topography as a function of spatial frequency.  {\bf B:} Number of SF components to reach 99$\%$ of the energy for each source. {\bf Left:} The component count visualised on orthogonal slices of brain volume. {\bf Right:} The 99$\%$-energy component count as a function of source depth. The dot correspond to individual sources and the lines estimate the medians of the distributions as a function of source depth.}
\label{fig:spatfreqsources}
\end{figure*}

\paragraph{Results} Fig. \ref{fig:spatfreqsources}A shows example SF representations of topographies with an increasing number of components. The energy of the on-scalp MEG topography is distributed to higher spatial frequencies than the energies of off-scalp MEG or EEG topographies. The distribution of energy also results in larger 99\% bandwidth than in the two other modalities.

Fig. \ref{fig:spatfreqsources}B displays the component count for 99$\%$ topography energy as a function of source location. Generally, this number decreases as a function of source depth. For the most superficial sources, the number of components ranges from 200 to 280 in on-scalp MEG, while in off-scalp MEG and in EEG it is less than $\sim$90. The 99$\%$ bandwidths of the most superficial sources in on-scalp MEG are about 210 1/m while in off-scalp MEG and EEG they are approximately 100 and 120 1/m, respectively. For the cortically-constrained source space, the 99$\%$ component count is about 300, 100 and 110 in on-scalp MEG, off-scalp MEG and EEG (bandwidths of 220, 100 and 130 1/m). For these values, the estimated sensor spacings were 1.5, 3.4 and 2.6 cm, matching a proportionality constant of 3.4 in the generalized sampling theorem described in Sec.~\ref{sec:sampling_grid_construction}.

\begin{figure*}[tbp]
\centering
\includegraphics[width=1\linewidth]{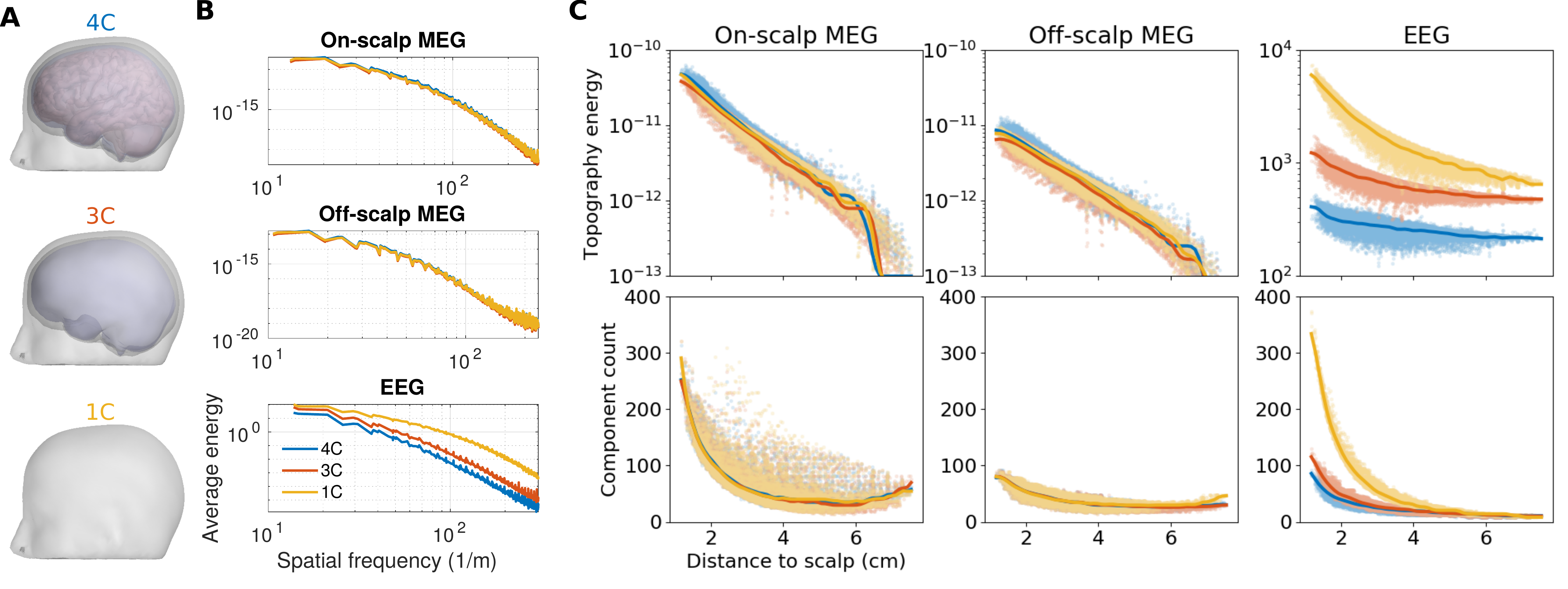}
\caption{The SF spectra and field energy with different volume conductors. {\bf A:} Volume conductors with different number of compartments (4C, 3C and 1C). {\bf B:} The topography energy averaged across the sources for different conductors. {\bf C:} Topography energy and 99$\%$ component count as a function of source depth.}
\label{fig:spatfreqheadmodel}
\end{figure*}

The effect of volume conduction on the topography energy spectra is visualized in Fig. \ref{fig:spatfreqheadmodel}. In on-scalp and off-scalp MEG, the energy spectra averaged over the sources shown in panel B is relatively unaffected by level of detail of the head model. By contrast, the average spectrum of EEG shows a strong dependence on the volume conduction. Comparing the spectrum of EEG computed using 4C to that using 1C, especially the energy in high spatial frequencies is reduced. In 1C, the component count for 99$\%$ energy is 300--350 for superficial sources in EEG, giving 99$\%$ bandwidth of about 250 1/m. The topography energy in on-scalp and off-scalp MEG is also only slightly affected by the head model: the energy of superficial sources increases with 4C. In contrast, the topography energy is reduced in EEG from 1C to 4C.

\subsection{Analysis of field kernel}

\paragraph{Methods} We analyzed the field kernel of identically and independently distributed (IID) random Gaussian sources (see Sec. \ref{sec:bioemsampling}). The sources were cortically constrained (cortically-constrained source space) and the effect of volume conduction was calculated using the 4C model. The 
eigendecomposition of the field kernel was computed by discretizing the kernel on the surface and solving discrete eigenvectors as in Eq.~\eqref{eq:disclb}. The field kernels were quantified by calculating the variance $K(\vec{r}, \vec{r})$ on the measurement surface. In addition, the correlation length of $K(\vec{r}_i, \vec{r})$ was quantified by computing the distance along the surface at which the kernel had decayed to half (the 'half-maximum width') for each measurement point $\vec{r}_i$. The distances along the surfaces were computed using a method based on the heat equation \cite{Crane2017HMD} implemented as the MATLAB function \texttt{heat\_geodesic} in gptoolbox \cite{gptoolbox}. 

The computation of the 99$\%$-energy component count was repeated using the field eigenbasis (Sec. \ref{sec:topoenergy}). The ratio of the 99$\%$ component counts in the eigenbasis and the SF basis was calculated to measure the compression in the topography representations.

\begin{figure*}[!t]
\centering
\includegraphics[width=1.0\linewidth]{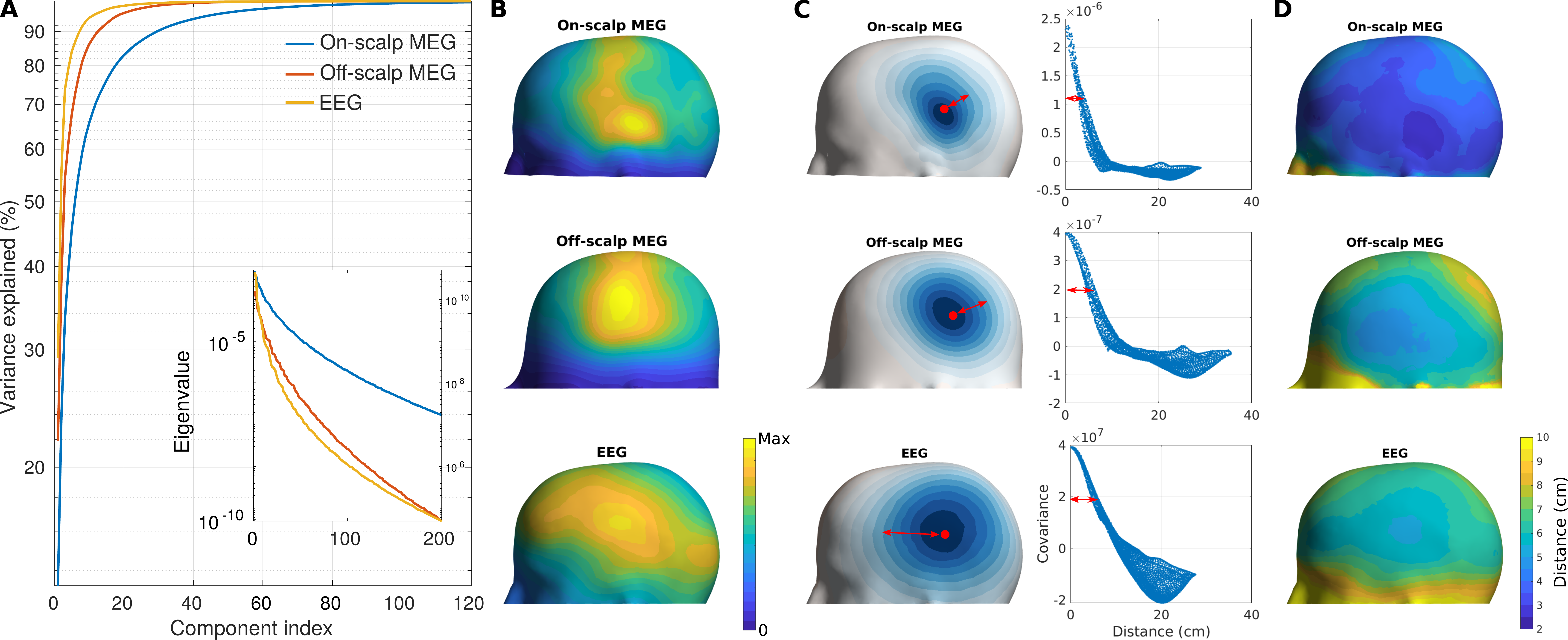}
\caption{Field kernel due to independent Gaussian sources with equal variance. {\bf A:} Field variance as a function of the field eigencomponent. Inset shows the eigenvalues. {\bf B:} Field variance on the measurement surface. {\bf C:} Spatial profile of the field kernel at one position on the measurement surface and its decay as a function of distance along the surface. The length at which the kernel has decayed to half (half-maximum width) is illustrated. {\bf D:} Half-maximum width of the field kernel across the measurement surface.}
\label{fig:fieldcovariance}
\end{figure*}

\paragraph{Results} Fig. \ref{fig:fieldcovariance} quantifies the field kernels of the IID random source distribution for the three modalities. When decomposed into eigencomponents, the cumulative variance grows the quickest in EEG and the slowest in on-scalp MEG. To explain 99$\%$ of the variance, 88, 35 and 25 eigencomponents are needed in on-scalp MEG, off-scalp MEG and EEG, respectively. Variance is distributed nonuniformly on the measurement surfaces with highest values around the temporal cortex. The kernels are asymmetric and the correlation length varies across the surface; the least correlated area can be found on top of the temporal cortex. The correlation lengths are shortest in on-scalp MEG as indicated by smaller values of half-maximum widths (average: 4.3 cm) compared to off-scalp MEG (7.2 cm) and EEG (7.2 cm).

\begin{figure}[!t]
\centering
\includegraphics[width=0.75\linewidth]{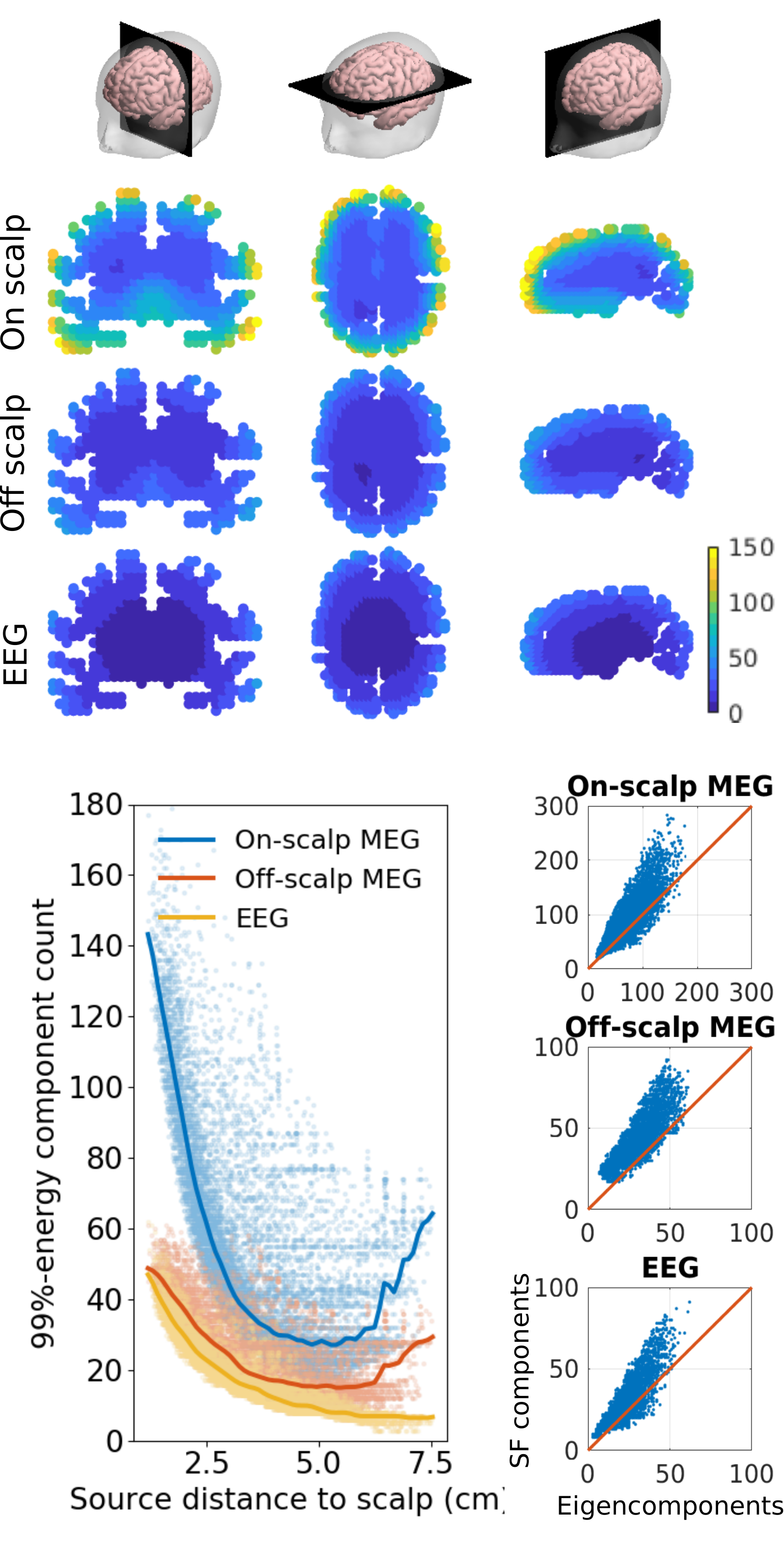}
\caption{The number of field eigencomponents that yield 99$\%$ of the topography energies as in Fig. \ref{fig:spatfreqsources}. {\bf Top:} The number of components required for 99$\%$ energy for each topography. {\bf Bottom left:} 99$\%$ eigencomponent count as a function of source depth. {\bf Bottom right:} Scatterplots across the source positions comparing the number of field eigencomponents and spatial-frequency (SF) components to reach 99$\%$ topography energy.}
\label{fig:compression}
\end{figure}

The number of field eigencomponents to achieve 99$\%$ topography energy (Eq.~\eqref{eq:cumen}) are shown in Fig. \ref{fig:compression}. The number of eigencomponents are at maximum 179, 61 and 62 for on-scalp MEG, off-scalp MEG and EEG. Overall, less eigencomponents are typically needed to reach 99$\%$ topography energy than SF components. The topography compression ratio is on average 81$\%$, 67$\%$ and 73$\%$ in on-scalp MEG, off-scalp MEG and EEG while the ranges are 42$\%$--200$\%$, 30$\%$-- 130$\%$ and 30$\%$--180$\%$, respectively.

\subsection{Sampling grid construction and evaluation}

\label{sec:methods_sampling_algorithm}

\paragraph{Method for grid construction} 
The sampling grids were constructed by subsampling the nodal positions of the triangle meshes using a farthest-point sampling algorithm implemented in gptoolbox \cite{gptoolbox}. As described in Sec. \ref{sec:sampling_grid_construction}, we maximized the total information based on embedding the sample positions in the eigenspace of the field kernel. With a given number of samples $N$, the number of components in the eigenspace was chosen larger than $N$. We found that the output of the optimization algorithm was highly sensitive to the initial configuration. Thereby, we ran the algorithm several times with random initial configurations to seek the global optimum.


\paragraph{Grid construction} We constructed sampling grids using the sampling algorithm described in Sec.~\ref{sec:sampling_grid_construction} for three different scenarios: a global scenario where the whole brain was assumed active and of interest, and two local scenarios where regions of interest (ROIs) in the brain were defined. We compared uniform sampling to model-informed sampling for different numbers of spatial samples $N$. The uniform sampling was generated by bandlimited priors in the SF basis meaning white covariance for the $N$ lowest SF coefficients (SF bandlimited prior; Sec. \ref{sec:bioemsampling}). Model-informed sampling was generated by encoding the prior in the covariance of the cortically-constrained sources (IID source prior; Sec. \ref{sec:bioemsampling}). The SF basis functions were computed using a zero-Dirichlet BC to avoid sampling near the boundary of the measurement surface. The grids were evaluated by computing FEV (Eq.~\eqref{eq:fev}) and information (Eq.~\eqref{eq:info}) from the 'ground-truth' model of the signal and noise.


\subsubsection{Global scenario}

\paragraph{Methods} In this scenario, the ground-truth model corresponded to IID source prior and white noise with standard deviation ranging from $0.05\sigma_0$ to $20\sigma_0$. The source standard deviation (7.7 pAm) in the ground-truth model was fixed so that the average SNR over the measurement surface was 1 for on-scalp MEG when the white noise level was $\sigma_0 = 9$  $\mathrm{fT}$. The spatial white noise levels $\sigma_0$ were then computed for off-scalp MEG (3.7 $\mathrm{fT}$) and EEG (42 $\mathrm{nV}$) using the same source variance and assuming that the average SNR was 1. The grids were constructed on meshes with the "eye areas" cropped as placing sample points on them was not desired.

\begin{figure*}[!t]
\centering
\includegraphics[width=1.0\linewidth]{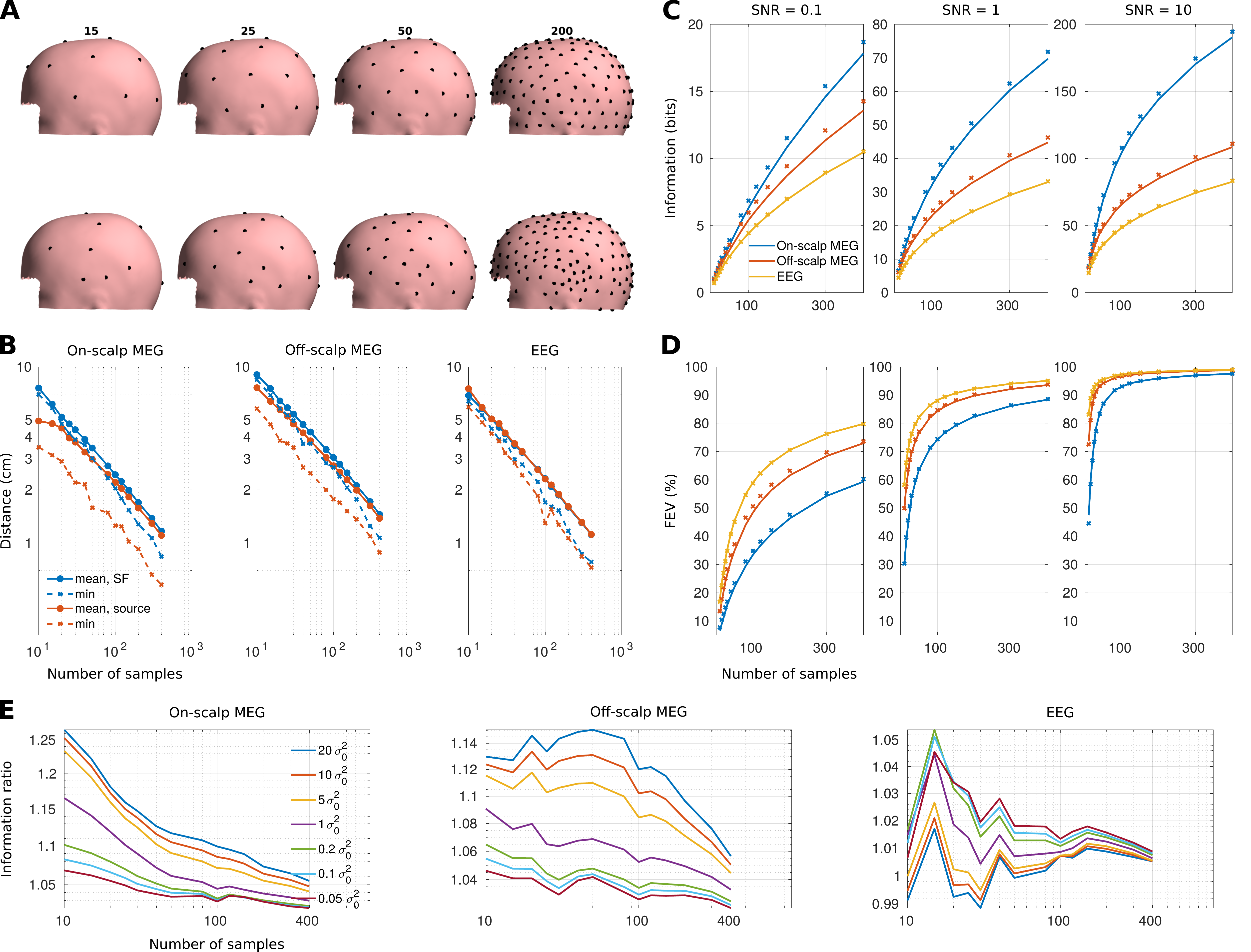}
\caption{Comparison of uniform (SF bandlimited prior) and model-informed (IID source prior) sampling of the whole brain in on-scalp MEG, off-scalp MEG and EEG. {\bf A:} Example on-scalp MEG arrays generated with the priors (top: SF; bottom: source). {\bf B:} Average and minimum nearest-pair distances of the sampling grids (blue: SF; red: source). {\bf C} and {\bf D:} Total information and fractional explained variance (FEV) computed with different SNRs (solid lines: SF; crosses: source). {\bf E:} Ratio of information obtained by model-informed and uniform sampling with different spatial white noise levels $\sigma_0^2$. }
\label{fig:iidsampling}
\end{figure*}

\paragraph{Results} Fig. \ref{fig:iidsampling} summarizes the sampling in the global scenario. For model-informed sampling, the sample density is the highest by the temporal lobe, which corresponds to the shorter correlation lengths and higher variance of the field kernel in Fig.~\ref{fig:fieldcovariance}. Model-informed sampling is beneficial in MEG at low SNR $<$ 1 and sample numbers giving about 15$\%$--25$\%$ increase in information compared to uniform sampling. The information gained from model-informed sampling decreases when SNR or sample number increases in on-scalp and off-scalp MEG. 

The total information is highest in on-scalp MEG, while the explained variance is largest in EEG. When the noise level is high (10$\sigma_0^2$), only 60$\%$, 74$\%$ and 80$\%$ of variance is explained (on-scalp MEG, off-scalp MEG and EEG, respectively) with sample number as high as 400. When the noise level is low (0.1$\sigma_0^2$), about 80, 30 and 20 samples are needed to explain 90$\%$ variance in on-scalp MEG, off-scalp MEG and EEG, respectively. Generally, as the number of samples increases the additional information provided by the sample decreases.

\begin{figure}[tbp]
\centering
\includegraphics[width=1.0\linewidth]{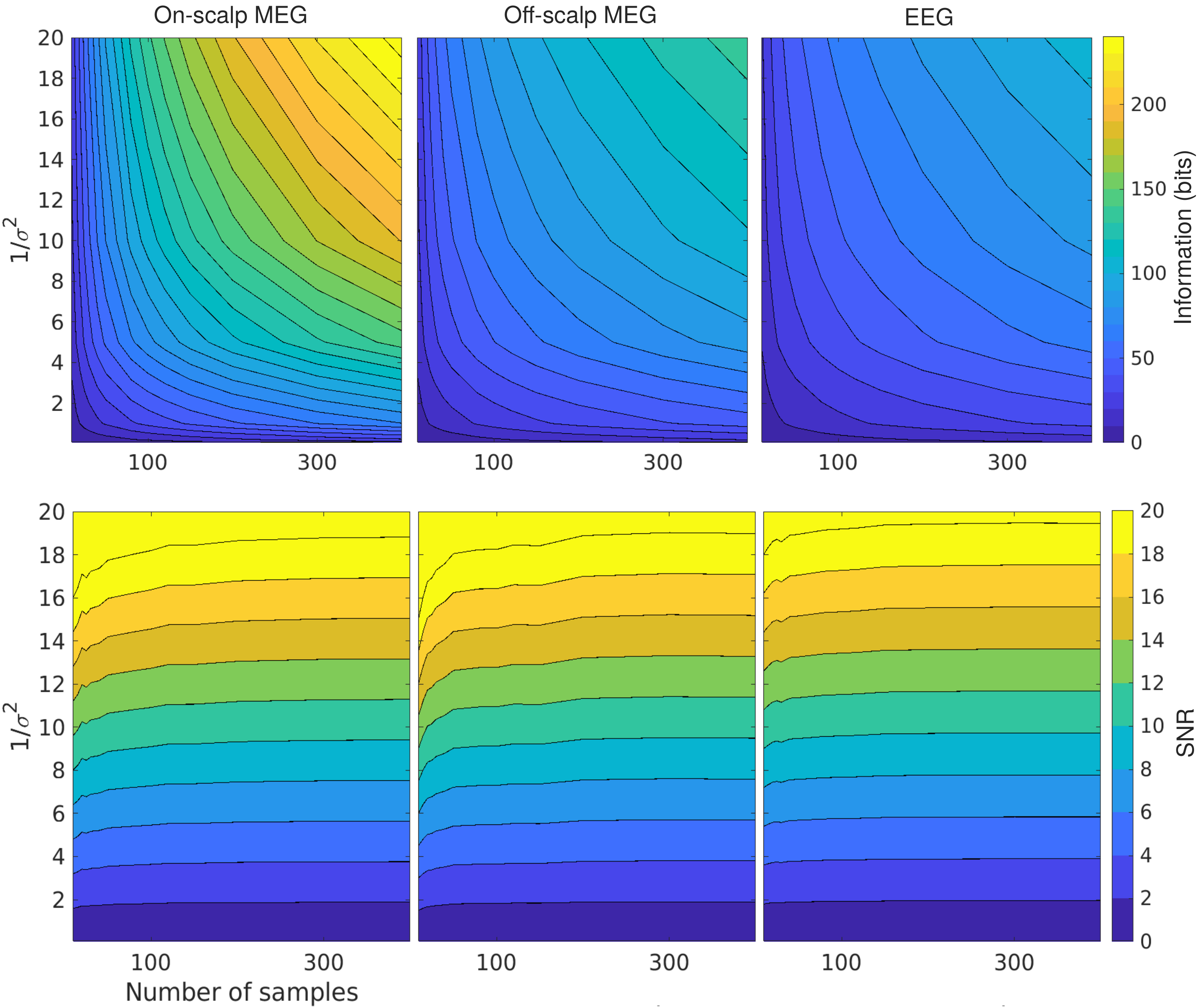}
\caption{Contourplots of total information (top) and SNR (bottom) as a function of inverse noise variance (1/$\sigma^2$; units 1/$\sigma_0^2$) and number of spatial samples for uniform sampling of the whole brain. The contour spacings are 10 bits and 2. The white noise variance $\sigma_0^2$ is different across the arrays (see the text).}
\label{fig:infocontours}
\end{figure}

The total information and SNR of uniform sampling grids as a function of noise variance and number of spatial samples are shown in Fig. \ref{fig:infocontours}. The SNR behaves similarly in different measurements and has a small dependence on the sample number. The total information increases more in on-scalp MEG than in off-scalp MEG or EEG, when noise variance is reduced or the number of samples is increased.

\subsubsection{Local scenarios}

\paragraph{Methods} In the local scenarios, the ground-truth models consisted of IID sources distributed in ROIs defined around the motor cortex. Both white (sensor) and colored (sensor + background brain activity) noise were considered. The ROIs consisted of sources within a patch that had either a diameter about 3 or 10 cm along the cortical surface (99 or 1 094 sources). Sampling grids were constructed for on-scalp MEG and EEG.

The source standard deviation $q$ (60 pAm) was fixed so that the maximum spatial SNR [max(SNR($\vec{r}$))] of the focal source distribution on the on-scalp MEG surface was 10 when the spatial white noise level was 9 $\mathrm{fT}$. The spatial white noise level of EEG (20 $\mathrm{nV}$) was fixed so that the source distribution gave also maximum SNR of 10. In addition to the white noise, colored noise due to brain background activity was included: the sources outside the ROI were assumed to generate field noise by modeling them as IID Gaussian with variance $q^2/100$.

The field kernel was whitened (Sec. \ref{sec:optimalcriteria}) and the spatial SNR was analyzed. The eigendecomposition of the whitened kernel was inspected to analyze the components contributing the most to the total information (\ref{sec:apppendix_info}). The SF basis for uniform sampling was constrained to a region on the surface with a distance to the center of the ROI less than 6 and 9 cm in focal and spread source distributions, respectively. For model-informed sampling, a field kernel generated with IID random sources in the ROI was used while noise kernel corresponded to either to white or colored noise models.

\begin{figure*}[tbp]
\centering
\includegraphics[width=1.0\linewidth]{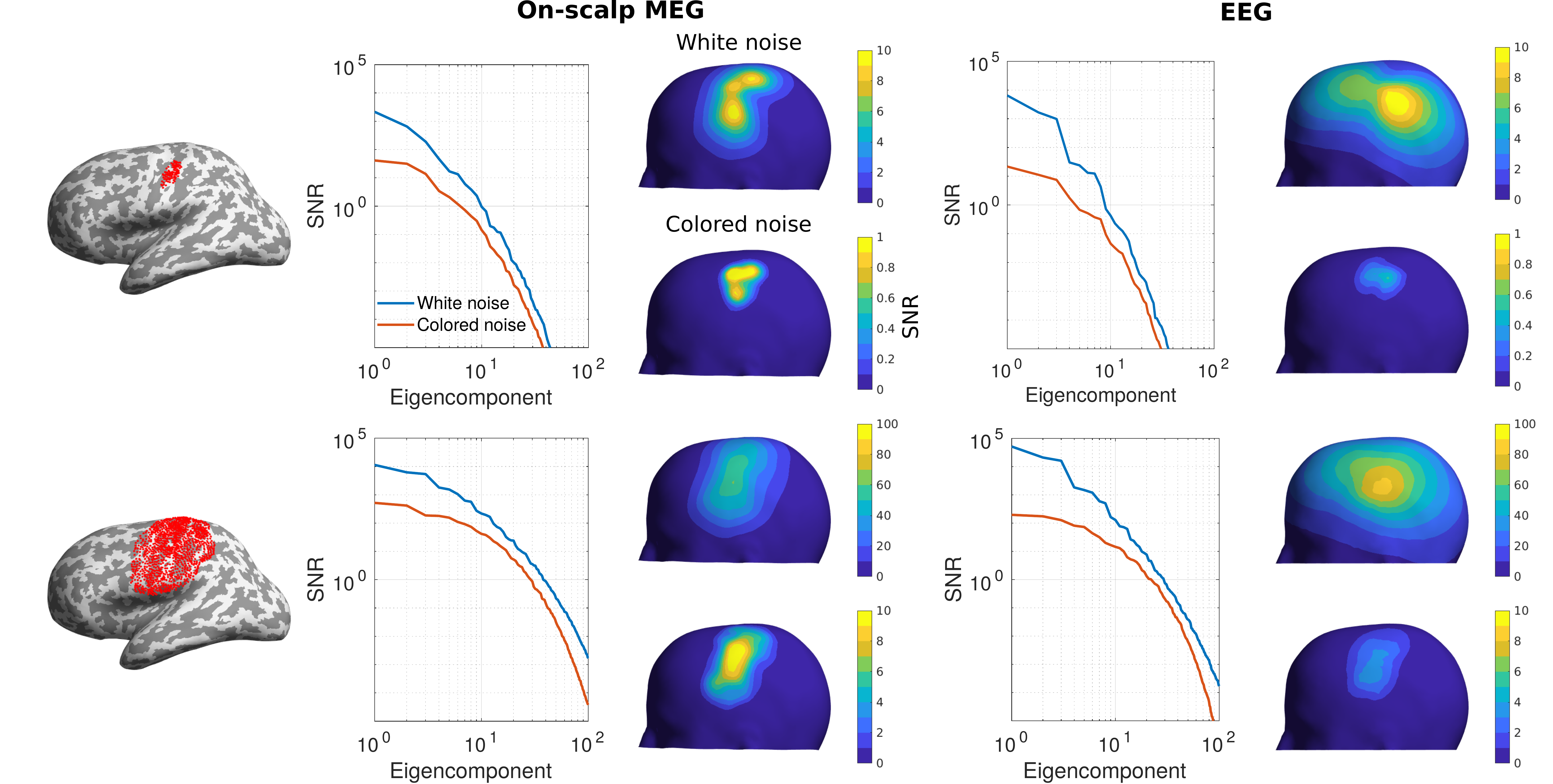}
\caption{Spatial and eigencomponent SNR distributions in on-scalp MEG and EEG due to sources in two regions of interest in the cortex with different size. SNRs for two noise models are shown: spatial white noise as well as colored noise due to brain background activity together with white noise. }
\label{fig:roistats}
\end{figure*}

\paragraph{Results} Fig. \ref{fig:roistats} shows the SNR distributions of on-scalp MEG and EEG for the two ROIs in the motor cortex. For the smaller ROI, the maximum spatial SNR across the whole measurement surface was 10 for both on-scalp MEG and EEG when only white noise was considered, while the maximum SNR of the eigencomponents were 2 100 (5.5 bits) and 6 500 (6.3 bits), respectively. When colored brain background noise was added, the maximum spatial SNR was reduced to 1.4 and 0.5, while maximum component-wise SNR was reduced to 41 (2.7 bits) and 22 (2.3 bits). Nine and eight components had SNR $>$ 1 in on-scalp MEG and EEG when white noise was considered, while the number of those components were six and four with colored noise. For the larger ROI, the maximum spatial SNRs were 60 and 84 with white noise and 10 and 4 with colored noise. The maximum component-wise SNRs were 11 000 (6.7 bits) and 51 000 (7.8 bits) with white noise and 500 (4.5 bits) and 200 (3.8 bits) with colored noise. In on-scalp MEG, 37 components had SNR $>$ 1 with white noise while 29 components had SNR $>$ 1 with colored noise. In EEG, the number of those components were 29 and 22, respectively.

\begin{figure*}[!t]
\centering
\includegraphics[width=1.0\linewidth]{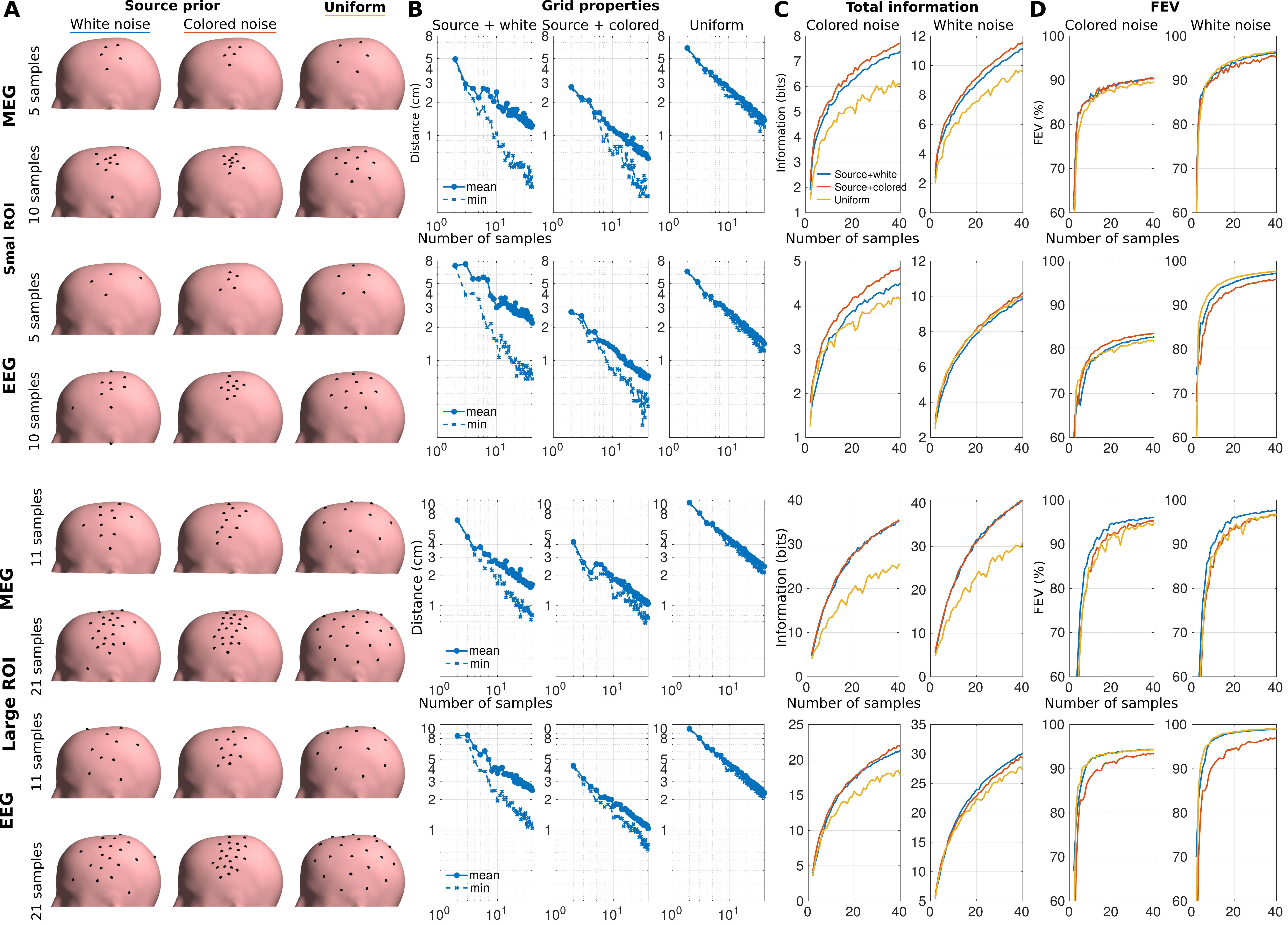}
\caption{Sampling of on-scalp MEG and EEG due to uncorrelated sources in an ROI around motor cortex. Two ROIs with diameters of roughly 3 cm (top) and 10 cm (bottom) are considered. {\bf A:} Example grids constructed with IID source and bandlimited SF ($\sim$uniform sampling) priors. {\bf B:} Average and minimum nearest-pair distances of the grids. {\bf C:} Information as a function of number of samples for the two noise models. {\bf D:} Fractional explained variance (FEV). }
\label{fig:roisampling}
\end{figure*}

The grids constructed using the different priors for signal and noise are presented in Fig. \ref{fig:roisampling}. Compared to uniform sampling, the model-informed grids are more densely distributed, especially when the colored noise model is used. The sample spacing in the model-informed grids tends to be smaller in MEG than in EEG. The colored noise model gives the highest information when the ROI is small; when the ROI is large, the grids generated with white and colored noise models are more equal in terms of information. The dense grids generated with white and colored noise priors also perform better than uniform grids when only white noise is considered. 

\section{Discussion}


\noindent We presented a theoretical framework for analyzing spatial degrees of freedom and correlations of fields on a surface. Based on this framework, we derived a method that maximizes the total information using the given number of samples. We applied the framework to analyze spatial sampling of EEG as well as of on- and off-scalp MEG.

We analyzed topographies of individual sources and field kernels of random source distributions. The source topographies were decomposed into spatial-frequency (SF) components as well as into field eigencomponents in order to determine the number of samples needed to capture 99\% of the topography energy in each basis. Using different priors for the field, we generated sampling grids when the whole brain or a part of it was of interest.

\subsection{Field analysis}

To achieve 99\% of the topography energy of every source in the brain, approximately 300, 100 and 110 SF components were needed in on-scalp, off-scalp MEG and EEG, respectively. These numbers represent the maximum number of spatial degrees of freedom in all of the topographies. Additionally, they correspond to the number of uniform samples needed to achieve at most roughly 1\% reconstruction error in noiseless conditions. These results are in a good accordance with previously published results. For MEG, this corresponds to the "rule of thumb" presented by Ahonen et al. \citeyear{ahonen1993}: the sensor spacing should be approximately the distance of the sensors to the closest sources. For EEG, this is in line with the arguments by Srinivasan et al. \citeyear{srinivasan1998}: the sensor count should be roughly 120 for a typical adult head. 

The spatial bandwidths of the topographies did not depend on the level of detail of the volume conductor in MEG. In contrast, the bandwidths in EEG were affected by the head tissues, demonstrating the well-known spatial low-pass filtering of EEG \cite{srinivasan1996}. Therefore, the spatial resolution of MEG is limited by the measurement distance while that of EEG is limited by the tissue conductivity contrasts.

The eigenbasis of the field kernel (Eq. \eqref{eq:mercer}) was shown to reduce the component count needed to explain 99\% of the topography energies, compressing the topography representations on average by 19\%, 33\% and 27\% compared to the SF basis (on-scalp, off-scalp MEG, and EEG). As a consequence, a smaller number of samples is needed for the same reconstruction accuracy than with the SF basis. However, to actually achieve the same accuracy, the samples must be placed in an non-uniform grid taking account the spatially varying correlation structure of the field kernel. 

\subsection{Grid construction}

To generate sampling grids, we presented a method that utilizes prior information of signal and noise in the form of a covariance kernel. 
Assuming random spatial frequency coefficients up to some bandlimit (bandlimited SF prior), the kernel becomes isotropic and translation-invariant and the presented method yields uniform sampling grids. 
With this approach, uniform grids may be constructed also on complex surfaces for applications that require uniform sampling. To analyze MEG and EEG sampling, we constructed kernels from the topographies of IID random neural sources. The shape and size of this model-informed kernel depends on location, thus yielding nonuniform sampling grids.


Compared to whole-head uniform sampling, whole-head nonuniform model-informed sampling was slightly beneficial in MEG: when the number of samples was small and the noise level high, nonuniform sampling yielded 10\%--25\% more information than uniform sampling. When decreasing the noise level or increasing the number of samples, the performance of uniform and nonuniform sampling grids was similar. In EEG, the benefits due to nonuniform sampling were less clear. The nonuniform grids generated with the method may have been suboptimal for EEG: the method assumed positive field correlations while EEG has negative correlations at long distances (see Fig. \ref{fig:fieldcovariance}). Altogether, among the measurements, with similar SNR and the number of samples, on-scalp MEG yielded the most information and more measurements were needed to explain the same amount of variance.

When a specific region of interest was defined, local high-density (nonuniform) sampling was beneficial in terms of total information. Dense spatial sampling of the local spatial degrees of freedom yielded higher information than uniform sampling covering larger measurement area. Dense sampling is especially useful when colored noise fields (such as those generated by background brain activity) are present and the region of interest is small. Local dense arrays would be beneficial in applications where only a certain part of the brain is of interest and the number of sensors is limited (e.g. \citealt{iivanainen2019potential}) or brain--computer interfacing where simple measurement setups are desirable. 




\subsection{Future directions}

The sampling grid optimization presented here does not necessarily result in sensor arrays that can be readily implemented in practice. For example, the minimum distance between the sensors was not constrained, the sensor dimensions were ignored, and the arrays were designed for specific use cases only. Modeling sensor dimensions is important as they limit the minimum distance between the sensors. Additionally, the sensor noise level is proportional to the dimensions for many sensor types (e.g., \citealt{mitchell2019quantum, kemppainen1989}). The spatial integration within a sensor can be viewed as a spatial low-pass filter reducing aliasing from higher frequency components \cite{roth1989using}. Nevertheless, the low-pass effect is estimated to be small unless the sensors are much larger than the correlation length of the field. 

The optimization method at its current stage is useful in understanding a theoretically optimal sensor array for a specific question such as sampling a certain brain region. It can be used to aid in the practical sensor-array design to inform whether it is beneficial to decrease sensor noise, add sensors or switch from uniform to nonuniform sampling. The sensor dimensions could be included and the method could be used with a population of head models so that a "universal" sensor array with the best average performance could be designed.

Our analysis was restricted to scalar fields which is already sufficient for EEG. In contrast, magnetic field is a vector field, but we only analyzed the field normal component.
One motivation for this was that SQUID-based MEG devices have typically measured the normal component and, additionally, a recent study showed that the normal component has the highest total information \cite{iivanainen_measuring_2017}. However, the same study also noted that the different field components provide mutually non-independent information. The sensor orientation could be included in the optimization by applying vector basis functions to assemble the kernels.



Besides adequate sampling of the neural fields, sensor arrays should also sample the interference fields in order to disentangle them from the field of interest. This would require even more spatial samples. For that purpose, in MEG, multiple layers of sensors at different distances from the scalp could be beneficial \cite{nurminen2010improving, nurminen2013improving}. For optimization of such arrays, the basis functions should be extended to 3D and the signal and interference (noise) fields should be presented as kernels corresponding to the prior knowledge as well as possible. In spherical geometry, the spherical-harmonics expansion of the field provides a basis with a "built-in support" for representation of signal and interference fields \cite{taulu2005presentation}. Finally, to reject homogeneous interference fields in MEG, sensors configured as gradiometers are commonly used. Gradiometers have also implications on the spatial sampling of the field \cite{ahonen1993} not considered in this work. 



With the generalization of the SF basis to arbitrary measurement surfaces, the standard 1D signal processing methods such as filtering and interpolation should be straightforward to apply for spatial data provided by EEG and MEG. Spatial-frequency filtering should be useful in noise and interference rejection as demonstrated already in \cite{graichen2015sphara}. The continuous SF basis functions used in this work make the filtering less dependent on the sensor configuration. As the SF basis is both data- and model-independent, the related filtering methods may be useful in real-time applications such as brain--computer interfaces. Such filtering as part of preprocessing would be similar to the spline Laplacian used in EEG \cite{nunez2019multi} with flexibility on defining the filter coefficients. Filtering and interpolation may also help in data visualization. 



It is often also of interest to estimate the neural sources that could have generated the data. This is performed by solving an inverse problem \cite{sarvas1987basic}. Here, we did not explicitly consider source estimation. However, total information should be an estimator of the source estimation performance of the array as it quantifies the size (or the stability of inversion) of the measurement covariance matrix. In other terms, when the field kernel is generated from the brain source prior, total information maximization corresponds to minimization of the posterior entropy of the sources (Sec. \ref{sec:optimalcriteria}). Moreover, the Gaussian random-field model has a direct connection to the minimum-norm source estimation (\ref{sec:appendix_mne}).



\section{Conclusions}

\noindent We presented a novel theoretical framework for spatial sampling of EEG and MEG. This framework bridges from the classic Shannon sampling and information theory to Bayesian experimental design and gives means to design optimal sensor arrays. Compared to off-scalp MEG and EEG, on-scalp MEG generally benefits from three times more spatial samples. When a certain region in the brain is of interest, high-density spatial sampling is beneficial so that sampling can be concentrated on a specific area in the brain.

\section*{Acknowledgments}
\noindent This work has received funding from the European Union’s Horizon 2020 research and innovation programme under grant agreement No 686865 (project BREAKBEN), the European Research Council under grant agreement No 678578 (project HRMEG), the National Institute of Neurological Disorders and Stroke of the National Institutes of Health under Award Number R01NS094604m, the Finnish Cultural Foundation under grant nos. 00170330 and 00180388 (JI), and Vilho, Yrjö and Kalle Väisälä Foundation (AM). The content is solely the responsibility of the authors and does not necessarily represent the official views of the funding organizations.

\section*{References}

\bibliographystyle{apa}
\bibliography{refs.bib}

\appendix
\section{Interpretations of posterior variables}
\label{sec:appendix_mne}
\noindent As the field model is linear, the field covariance obeys the form $K(\vec{r}, \vec{r}\,') = \mathbf{\psi}(\vec r)^\top\mathbf{K}_a\mathbf{\psi}(\vec r')$. As $\mathbf{k}(\vec{r}) = \boldsymbol{\Psi}\mathbf{K}_a\mathbf{\psi}(\vec{r})$, we can read from Eq.~\eqref{eq:posterior_bayes} that the posterior covariance of the coefficients $\mathbf{a}$ is
\begin{equation}
    \mathbf{K}_a^* = \mathbf{K}_a - \mathbf{K}_a\boldsymbol{\Psi}^\top
    (\mathbf{K} + \boldsymbol{\Sigma})^{-1}
    \boldsymbol{\Psi} \mathbf{K}_a.
\end{equation}
Direct application of Woodbury matrix identity \cite{petersen2008matrix} yields
\begin{equation}
\begin{split}
    \mathbf{K}_a^*
        &= \mathbf{K}_a - \mathbf{K}_a\boldsymbol{\Psi}^\top
    (\mathbf{K} + \boldsymbol{\Sigma})^{-1}
    \boldsymbol{\Psi} \mathbf{K}_a
    \\ &= \mathbf{K}_a - \mathbf{K}_a\boldsymbol{\Psi}^\top
    (\boldsymbol{\Psi} \mathbf{K}_a \boldsymbol{\Psi^\top} + \boldsymbol{\Sigma})^{-1}
    \boldsymbol{\Psi} \mathbf{K}_a
    \\ &=(\boldsymbol{\Psi}^\top \boldsymbol{\Sigma}^{-1} \boldsymbol{\Psi} + \mathbf{K}_a^{-1})^{-1},
\end{split}
\end{equation}
which becomes the classical covariance of the coefficient estimates of Eq.~\eqref{eq:coef_covariance}, when $\mathbf{K}_a^{-1}$ approaches zero and the measurement noise is white. 

Similarly, the posterior mean of the coefficients is 
\begin{equation}
\begin{split}
    \boldsymbol{\mu}_a^* &= \mathbf{K}_a\boldsymbol{\Psi}^\top (\mathbf{K} + \boldsymbol{\Sigma})^{-1}\mathbf{y} \\
    &=\mathbf{K}_a\boldsymbol{\Psi}^\top (\boldsymbol{\Psi \mathbf{K}_a \Psi^\top} + \boldsymbol{\Sigma})^{-1}\mathbf{y},
\end{split}    
\end{equation}
which follows the form of the generalized minimum-norm estimate \cite{dale1993improved} well known in EEG/MEG. Inserting $\boldsymbol{\Sigma} = \sigma^2\mathbf{I}$ and $\mathbf{K}_a = \lambda^2\mathbf{I}$, we get the coefficients to the form $\boldsymbol{\Psi}^\top(\boldsymbol{\Psi}\boldsymbol{\Psi}^\top  + (\sigma^2/\lambda^2)\mathbf{I})^{-1}\mathbf{y}$, which is the more common form for the estimator, $\sigma^2/\lambda^2$ being the Tikhonov regularization parameter.

The coefficient estimator can be further manipulated:
\begin{equation}
    \begin{split}
    \boldsymbol{\mu}_a^* &= \mathbf{K}_a\boldsymbol{\Psi}^\top (\mathbf{K} + \boldsymbol{\Sigma})^{-1}\mathbf{y}
    \\ &= \mathbf{K}_a\boldsymbol{\Psi}^\top (\mathbf{K} + \boldsymbol{\Sigma})^{-1}[(\mathbf{K} + \boldsymbol{\Sigma}) - \mathbf{K}] \boldsymbol{\Sigma}^{-1}\mathbf{y} \\
    &= \mathbf{K}_a\boldsymbol{\Psi}^\top [\mathbf{I} - (\mathbf{K} + \boldsymbol{\Sigma})^{-1}\mathbf{K}] \boldsymbol{\Sigma}^{-1}\mathbf{y} \\
    &= [\mathbf{K}_a\boldsymbol{\Psi}^\top - \mathbf{K}_a\boldsymbol{\Psi}^\top(\mathbf{K} + \boldsymbol{\Sigma})^{-1}\boldsymbol{\Psi}\mathbf{K}_a\boldsymbol{\Psi}^\top] \boldsymbol{\Sigma}^{-1}\mathbf{y} \\
    &= [\mathbf{K}_a - \mathbf{K}_a\boldsymbol{\Psi}^\top(\mathbf{K} + \boldsymbol{\Sigma})^{-1}\boldsymbol{\Psi}\mathbf{K}_a] \boldsymbol{\Psi}^\top\boldsymbol{\Sigma}^{-1}\mathbf{y} = \mathbf{K}_a^*\boldsymbol{\Psi}^\top\boldsymbol{\Sigma}^{-1}\mathbf{y} \\
    &= (\boldsymbol{\Psi}^\top \boldsymbol{\Sigma}^{-1} \boldsymbol{\Psi} + \mathbf{K}_a^{-1})^{-1}\boldsymbol{\Psi}^\top\boldsymbol{\Sigma}^{-1}\mathbf{y},
    \end{split}
\end{equation}
which again reduces to the least-squares estimator of Eq.~\eqref{eq:ord_least_squares}, when $\mathbf{K}_a^{-1}$ approaches zero and the measurement noise is white.

\section{Total information and covariance}
\label{sec:apppendix_info}
\noindent Channel capacity or the total information conveyed by a noisy channel \cite{shannon1949} has been used to evaluate different sensor arrays in MEG \cite{kemppainen1989, nenonen2004total,iivanainen_measuring_2017,riaz2017evaluation}. Here we show how the channel capacity relates to the sample and noise covariances, $\mathbf{K}$ and $\boldsymbol{\Sigma}$, described in Sec.~\ref{sec:bayes_sampling}.

If the field samples as well as the noise were uncorrelated, the total information would be $1/2\sum_i \log_2(P_i + 1)$ where $P_i$ is the power signal-to-noise ratio of each measurement. If the measurements are correlated, they can be orthogonalized by the eigendecomposition of whitened covariance matrix $\Tilde{\mathbf{K}} = \boldsymbol{\Sigma}^{-1/2}\mathbf{K}\boldsymbol{\Sigma}^{-\top/2} = \mathbf{V}\mathbf{P}\mathbf{V}^\top$, where $\mathbf{V}$ contains eigenvectors of $\Tilde{\mathbf{K}}$ and $\mathbf{P}$ is a diagonal matrix of $P_i$.
Starting from the original formula, the information can be now written as 
\begin{equation}
\begin{split}
\mathrm{info}(R) &= \frac{1}{2}\sum_i \log_2(P_i + 1) = \frac{1}{2} \log_2 \det(\mathbf{P} + \mathbf{I}) 
\\ &=\frac{1}{2} \log_2 \det(\tilde{\mathbf{K}} + \mathbf{I}) 
= \frac{1}{2} \log_2 \frac{\det(\mathbf{K} + \boldsymbol{\Sigma})}{\det(\boldsymbol{\Sigma})}.
\end{split}
\end{equation}

The determinant measures the "size" of $\mathbf{K} + \boldsymbol{\Sigma}$. The determinants can also be interpreted as volumes spanned by the possible measurements and noise in the $N$-dimensional signal space ($N$ being the number of channels). This leads us to Shannon's original geometric interpretation: information measures (logarithmically) how many distinct signals are there in the signal space when each signal is surrounded by a volume of uncertainty $\det(\boldsymbol{\Sigma})$.

Total information can be maximized by choosing the measurement grid so that $\tilde{\mathbf{K}}$ is diagonal, i.e., each sample measures independent information. This can be seen from the matrix derivative \cite{petersen2008matrix}
\begin{equation}
    \frac{\partial \ln\, \det(\mathbf{X} + \mathbf{I}) }{\partial \mathbf{X}} = 2(\mathbf{X} + \mathbf{I})^{-1} - \mathbf{I} \odot (\mathbf{X} + \mathbf{I})^{-1},
\end{equation}
where $\odot$ is element-wise product and element-wise product with identity results the diagonal-part of the matrix. The logarithm of determinant gets maximized when the derivative with respect to its elements is zero, which means that the non-diagonal elements of $(\mathbf{X} + \mathbf{I})^{-1}$ must be zero. This is equivalent of $\mathbf{X}$ being diagonal itself.

For diagonal elements no solution exist, but their derivatives approach zero as the elements themselves approach infinity. This makes sense because the total information grows logarithmically with the signal variance.

\end{document}